\documentclass[12pt]{article}
\usepackage{amsfonts}
\usepackage{amsmath}
\usepackage{amssymb}
\usepackage{amsthm}
\usepackage{accents}
\usepackage{youngtab}
\usepackage{braket}
\usepackage{graphicx}
\usepackage[svgnames]{xcolor}
\usepackage{ytableau,tikz}
\usepackage{here}
\usepackage{comment}
\usepackage{tikz}
\usepackage{feynmf}
\usepackage{chessfss}

\newcommand{\ba}{\begin{eqnarray}}
\newcommand{\ea}{\end{eqnarray}}
\newcommand{\nn}{\nonumber}

\newcommand{\cW}{\mathcal{W}}

\newcommand{\cN}{\mathcal{N}}
\newcommand{\lt}{\left}
\newcommand{\rt}{\right}
\newcommand{\cO}{\mathcal{O}}

\usepackage[colorlinks=true,linkcolor=black,citecolor=black,urlcolor=black]{hyperref}

\makeatletter
\@addtoreset{equation}{section}

\makeatother

\begin{document}

\begin{titlepage}
\vspace*{-2cm}
	\begin{flushright}
		DIAS-STP-19-07\\
		UT-19-21
	\end{flushright}
	
	\vskip 1cm
	
	\begin{center}
		\textbf{\LARGE Testing Macdonald Index as a Refined Character of Chiral Algebra}
	\end{center}
	\vskip 2.5cm
		\begin{center}
			{\Large Akimi Watanabe$^\WhiteBishopOnWhite$ and Rui-Dong Zhu$^\WhiteKnightOnWhite$}
			\\[.6cm]
			{\em  $^\WhiteBishopOnWhite$Department of Physics, The University of Tokyo}\\
		{\em Bunkyo-ku, Tokyo 113-8654, Japan}\\
		        {\em  $^\WhiteKnightOnWhite$School of Theoretical Physics, Dublin Institute for Advanced Studies}\\
			{\em  10 Burlington Road, Dublin, Ireland}
			\\[.4cm]
		\end{center}
	\vfill
	\begin{abstract}
	We test in $(A_{n-1},A_{m-1})$ Argyres-Douglas theories with ${\rm gcd}(n,m)=1$ the proposal of Song's in \cite{Song} that the Macdonald index gives a refined character of the dual chiral algebra. In particular, we extend the analysis to higher rank theories and Macdonald indices with surface operator, via the TQFT picture and Gaiotto-Rastelli-Razamat's Higgsing method. We establish the prescription for refined characters in higher rank minimal models from the dual $(A_{n-1},A_{m-1})$ theories in the large $m$ limit, and then provide evidence for Song's proposal to hold (at least) in some simple modules (including the vacuum module) at finite $m$. We also discuss some observed mismatch in our approach for surface operators with large vortex number. 
	\end{abstract}
	\vfill
\end{titlepage}

\noindent\hrulefill
\tableofcontents
\noindent\hrulefill

\section{Introduction}

The duality between 4d $\cN=2$ gauge theories and 2d TQFT discovered in \cite{GRRY1,GRRY2} provides an extremely convenient approach to the superconformal index and BPS spectrum of 4d $\cN=2$ theories. The key ingredient here is the class S contruction of 4d $\cN=2$ theories \cite{classS}, via the compactification of 6d $\cN=(2,0)$ theory with a Lie-algebraic label $G$ on a punctured Riemann surface. The superconformal index is rephrased as a correlator of the dual TQFT, where each puncture is mapped to a state in 2d. This duality in particular casts light onto theories with no Lagrangian description, such as Argyres-Douglas (AD) theories discovered in \cite{AD-original}. Argyres-Douglas theories are classified from the viewpoint of class S construction with irregular punctures in \cite{Xie,Wang-Xie}, and it was found that only a sphere with one irregular puncture and a sphere with one irregular and one regular puncture are allowed to be the Riemann surface used to construct an AD theory. The full superconformal index depends on three fugacity parameters $(p,q,t)$ associated to three charges in the superconformal algebra that commute with the Hamiltonian. The wavefunction of irregular punctures in the Schur ($p=0$, $q=t$) and Macdonald ($p=0$) limit is simplified a lot and has been studied in \cite{BN, Song-TQFT, BN2,BN3}, and we will investigate the Macdonald index with tools developed in these previous works. 

In the Macdonald limit, only a special set of BPS operators, called Schur operators, contribute to the superconformal index. Interestingly, it was found in \cite{ChiralSym} that Schur operators span a chiral algebra containing the Virasoro symmetry on a 2d plane inside 4d $\cN=2$ superconformal theories. The Schur index plays the role of the vacuum character of the chiral algebra. As it is the same set of operators contributing to the Macdonald index, it is natural to expect the Macdonald index to give a refined character of the chiral algebra. Indeed it was observed in \cite{Song} that the Macdonald index matches with the refined character with a refinement parameter $T$ counting the number of basic generators used to generate each state in the chiral algebra. For example, the dual chiral algebras of  $(A_1,A_{2k})$ theories are known to be Virasoro minimal models \cite{CS}, and we count the number of Virasoro generators in each basis of the Verma module with weight $T$ to compute the refined character.The relation between the Macdonald index and the refined character of the vacuum module has been further studied in \cite{ALS,Xie-Yan} from the viewpoint of the vertex operator algebra (VOA), and the refinement described above was found to be identified with the Kazhdan filtration in the context of VOA. Related recent development is also presented in \cite{VOA4dN4,freefieldhiggs,rank-2-VOA}, where the corresponding filtration, called $R$-filtration, is identified in some other more general cases via the free field realization in the VOAS approach. 

The duality also holds for non-vacuum modules of the chiral algebra. It was found in \cite{CGS,CGS-surface,CGS-chiral} that defect operators play the role of (linear combinations of) primary operators in the chiral algebra. In particular, the Schur index with surface operator with label $\{s_i\}$ gives the character of the module labeled by the same set of parameters $\{s_i\}$ in a clean way in Argyres-Douglas theories \cite{NSZ}. The bootstrap of surface operators in the Schur limit has also been studied in \cite{bootstrap-surface}. We would like to try to compute the Macdonald index with surface operator for $(A_{n-1},A_{m-1})$ AD theories with ${\rm gcd}(n,m)=1$ in this article, and compare our results with the corresponding refined character. The tool we will use in the gauge theory side is the TQFT picture of AD theories and the Higgsing method that generates a surface operator extending in the orthogonal directions to the chiral algebra plane, developed in \cite{Higgsing} and rewritten in terms of the class S construction in \cite{NSZ}. 

The article is organized in the following way. We review some basic definitions and facts on superconformal index in section \ref{s:index}. We describe in section \ref{s:HL-wave} and \ref{s:Mac-wave} the way to extract out the wavefunction of irregular punctures $I_{n,m}$ from the regular puncture in the Hall-Littlewood and Macdonald limit and present several explicit expressions for $n=3$ by using the properties of $A_2$ Macdonald polynomials reviewed in Appendix \ref{a:Macdonald}. In section \ref{s:Higgsing}, we give the details on the Higgsing method we use to insert surface operators in AD theories, and we list the expressions of Macdonald indices thus obtained in section \ref{s:surface}. Song's prescription for refined character is described in section \ref{s:Song} and we provide an alternative approach, the POSET method, to the refined character in the same section. In section \ref{s:dis}, we compare the Macdonald index and the corresponding refined character, and also give the reasoning of our conjecture on wavefunctions presented in section \ref{s:Mac-wave}. We give a brief discussion about the fusion rules of surface operators in section \ref{s:fusion} before we conclude the article. 

\section{The Superconformal Index, TQFT picture and its Special Limits}\label{s:index}

The 4d $\cN=2$ superconformal index introduced in \cite{Aharony:2003sx,Kinney:2005ej} is defined as a Witten index,  
\ba
{\cal I}(p,q,t;\xi)={\rm tr}\lt[(-1)^F p^{\frac{E-2j_1-2R-r}{2}}q^{\frac{E+2j_1-2R-r}{2}}t^{R+r}\xi^fe^{-\beta H}\rt],\label{index-pqt}
\ea
in terms of the representation theory data $(E,j_1,j_2,R,r)$ of the $\cN=2$ superconformal group SU($2,2|2$) and $f$ of the flavor symmetry group. The Hamiltonian here is chosen\footnote{The superconformal index is known to be independent of this choice. For more details on the superconformal algebra, refer to for example \cite{ChiralSym}. } to be 
\ba
H=2\lt\{\bar{\cal Q}_{1\dot{-}},\bar{\cal Q}^\dagger_{1\dot{-}}\rt\}=2(E-2j_2-2R+r),
\ea
and as a well-known property for the Witten index, the superconformal index does not depend on $\beta$, and only counts BPS states satisfying $H=0$ or equivalently 
\ba
E=2j_2+2R-r.
\ea
We adopted the $(p,q,t)$ basis for the fugacity parameters in the superconformal index, which will be convenient in this article. 

A dual description of the superconformal index of class S theories by TQFT on the corresponding Riemann surface was conjectured in \cite{GRRY1}, that is 
\ba
{\cal I}(p,q,t;\xi)=\sum_\lambda \lt(C_\lambda(p,q,t)\rt)^{2g-2+m}\prod_{i=1}^m f^{(i)}_\lambda(p,q,t;\xi^{(i)}),
\ea
where a special basis that diagonalizes the three-point structure coefficient $C_\lambda$ is taken, $g$ denotes the genus of the Riemann surface that characterizes the class S theory, and $m$ denotes the number of punctures with $f^{(i)}_\lambda$ represents the corresponding wavefunction of the $i$-th puncture in TQFT. In particular, in the so-called Macdonald limit $p\rightarrow 0$, the diagonal basis is shown to be spanned by Macdonald polynomials \cite{GRRY2}, and the wavefunction for the regular puncture is also written in terms of the Macdonald polynomials. As one can see from the definition of the superconformal index (\ref{index-pqt}), non-trivial contributions to the superconformal index in the Macdonald limit has to satisfy one more constraint in addition to $H=0$, 
\ba
E-2j_1-2R-r=0.
\ea
These two conditions, which can be summarized into 
\ba
E=j_1+j_2+2R,\quad r=j_2-j_1,
\ea
define the notion of Schur operators. The Schur operators are classified through the representation theory of the superconformal algebra \cite{ChiralSym}, and this restricted spectrum simplifies the computation of the index a lot. We will focus on the computation of the Macdonald index, i.e. the Macdonald limit of the superconformal index, of $(A_{n-1},A_{m-1})$ Argyres-Douglas theories, and these theories can be constructed in the class S way by putting one irregular puncture on the sphere. The Macdonald index for these theories can thus be simplified to 
\ba
{\cal I}_{(A_{n-1},A_{m-1})}(q,t;\xi)=\sum_\lambda \lt(C_\lambda(q,t)\rt)^{-1}f^{I_{n,m}}_\lambda(q,t;\xi),
\ea
and when the theory has no flavor symmetry, we can further suppress the fugacity $\xi$. The wavefunction $f^{I_{n,m}}_\lambda$ was worked out for the series ${\rm gcd}(n,m)=n$ in \cite{BN,BN2,BN3}, and for the series ${\rm gcd}(n,m)=1$ in \cite{Song-TQFT}. Some details on the wavefunction necessary for this article will be reviewed in section \ref{s:HL-wave} and \ref{s:Mac-wave}. We will try to determine the wavefunction in the Macdonald limit for $I_{n,m}$ with ${\rm gcd}(n,m)=1$ in higher rank cases by mimicking the computation done in \cite{Song-TQFT}, and it involves a guessing work to interpolate between the so-called Hall-Littlewood limit, which is achieved by further taking $q\rightarrow 0$ in the Macdonald limit, and the so-called Schur limit, which specializes $q=t$ in the Macdonald limit. The Hall-Littlewood and the Schur limit are interesting in their own ways, as they are respectively related to the Coulomb branch index of 3d mirror \cite{3dmirror} and the 2d chiral algebra \cite{ChiralSym}. 

\section{Wavefunction in Hall-Littlewood Limit}\label{s:HL-wave}

The wavefunction for the (full) regular puncture in the Hall-Littlewood (HL) limit is given by\footnote{Note that $z_i$'s here are the root parameters, and the variables $x_i$'s are related to $z$ through, $z^{\alpha_{ij}}=x_i/x_j$ for the root $\alpha_{ij}=e_i-e_j$. We have for example for $n=3$, $x_1=z_1$, $x_2=z_2/z_1$, and thus the highest weight of the representation associated to the Young diagram $(\lambda_1,\lambda_2)$ is $\vec{w}=(\lambda_1-\lambda_2,\lambda_2)$.} 
\ba
f^{HL}_\lambda(z)=\frac{\tilde{P}^{HL}_\lambda(z)}{(1-t)^r\prod_{\alpha\in\Delta}(1-tz^\alpha)},
\ea
where from the normalization of the wavefunction, 
\ba
\frac{1}{n!}\oint\prod_i \frac{{\rm d}z_i}{2\pi iz_i}\prod_{\alpha\in\Delta}(1-z^\alpha)I^{HL}_{vec}(z)f^{HL}_{\lambda}(z)f^{HL}_{\mu}(z^{-1})=\delta_{\lambda\mu},
\ea
with 
\ba
I^{HL}_{vec}(z)=(1-t)^r\prod_{\alpha\in\Delta}(1-tz^\alpha),
\ea
we see that the normalized Hall-Littlewood polynomials satisfy 
\ba
\frac{1}{(1-t)^r} \oint\prod_i \frac{{\rm d}z_i}{2\pi iz_i}\Delta_t(z)\tilde{P}^{HL}_\lambda(z)\tilde{P}^{HL}_\mu(z^{-1})=\delta_{\lambda\mu}.
\ea
Therefore, the normalized Hall-Littlewood polynomial is given by 
\ba
\tilde{P}^{HL}_\lambda(z)=(1-t)^{\frac{r}{2}} \bar{P}^{HL}_\lambda(z|t),
\ea
where the expression and properties of $\bar{P}^{HL}_\lambda(z|t)$ is reviewed in Appendix \ref{a:HL}, and $r=n-1$ is the rank of the associated Lie algebra of the Hall-Littlewood polynomial, or equivalently the rank of the gauge group label $G$ of the AD theory in the class S picture. 

The wavefunctions of the irregular singularities $I_{n,-n+1}$ and $I_{n,0}$, as they can be used to construct certain (non-conformal) Lagrangian theories, can be worked out\footnote{This procedure does not only apply in some special limits, but also in the general full index case, as long as the TQFT picture holds. We only present it in the Macdonald limit here for relevance.} through gauging the flavor symmetry attached to the full regular puncture used in an equivalent construction \cite{Song-TQFT}, 
\ba
&&f^{I_{n,-n+1}}_\lambda(q,t)=\frac{1}{n!}\oint\prod_i \frac{{\rm d}z_i}{2\pi iz_i}\prod_{\alpha\in\Delta}(1-z^\alpha)I_{vec}(z)f_{\lambda}(z),\\
&&f^{I_{n,0}}_\lambda(a;q,t)=\frac{1}{n!}\oint\prod_i \frac{{\rm d}z_i}{2\pi iz_i}\prod_{\alpha\in\Delta}(1-z^\alpha)I_{vec}(z)I_{hyp}(z,a;q,t)f_{\lambda}(z).
\ea

For $n=2$, the wavefunction for the irregular puncture $I_{2,-1}$ can be found through\footnote{Note that for $n=2$, we always have $\bar{P}^{HL}_\lambda(z|t)=\bar{P}^{HL}_\lambda(z^{-1}|t)$.}
\ba
f^{I_{2,-1}}_\lambda(t)=\frac{1}{2}\oint\frac{{\rm d}\zeta}{2\pi i\zeta}(1-\zeta^2)(1-\zeta^{-2})I^{HL}_{vec}(\zeta;t)f^{HL}_\lambda(\zeta)=\oint\frac{{\rm d}\zeta}{2\pi i\zeta}\Delta_t(\zeta)(1-t\zeta^2)(1-t\zeta^{-2})\tilde{P}^{HL}_\lambda(\zeta)\nn\\
=\oint\frac{{\rm d}\zeta}{2\pi i\zeta}\Delta_t(\zeta)(\sqrt{1+t}\bar{P}^{HL}_\emptyset(\zeta|t)-t\bar{P}^{HL}_2(\zeta|t))\tilde{P}^{HL}_\lambda(\zeta)\nn\\
=\lt\{\begin{array}{cc}
\sqrt{1-t^2} & \lambda=0,\\
-t\sqrt{1-t} & \lambda=2,\\
0 & {\rm otherwise}.
\end{array}\rt.
\ea
For irregular singularities of type $I_{2,2m-1}$ with $m\geq1$, it is conjectured in \cite{Song-TQFT} that 
\ba
f^{I_{2,2m-1}}_\lambda(t)=\sqrt{1-t^2}\delta_{\lambda0}.
\ea
One can similarly compute the wavefunction for $I_{2,0}$ by evaluating the integral 
\ba
f^{I_{2,0}}_\lambda(a;t)=\frac{1}{2}\oint\frac{{\rm d}\zeta}{2\pi i\zeta}(1-\zeta^2)(1-\zeta^{-2})I^{HL}_{vec}(\zeta;t)I^{HL}_{hyp}(\zeta,a;t)f^{HL}_\lambda(\zeta),
\ea
where 
\ba
I^{HL}_{hyp}(\zeta,a;t)&=&\frac{1}{(1-t^{\frac{1}{2}}\zeta^\pm a^\pm)}.
\ea
As we have  
\ba
&&\lt((1+t)\chi^{su(2)}_0(\zeta)+t^2\chi^{su(2)}_0(\zeta)-t\chi^{su(2)}_2(\zeta)\rt)I^{HL}_{hyp}(\zeta,a;t)\nn\\
&&=\frac{\sqrt{1+t}}{1-t}\bar{P}^{HL}_\emptyset(\zeta|t)+\frac{1}{1-t}t^{\frac{1}{2}}(a+a^{-1})\bar{P}^{HL}_1(\zeta|t)+\frac{1}{1-t}t(a^2+a^{-2})\bar{P}^{HL}_2(\zeta|t)\nn\\
&&+\frac{1}{1-t}t^{\frac{3}{2}}(a^3+a^{-3})\bar{P}^{HL}_3(\zeta|t)+\frac{1}{1-t}t^{2}(a^4+a^{-4})\bar{P}^{HL}_4(\zeta|t)+\dots,
\ea
we obtain 
\ba
f^{I_{2,0}}_\lambda(a;t)=\lt\{\begin{array}{cc}
\sqrt{\frac{1+t}{1-t}} & \lambda=0,\\
\frac{t^{\frac{\lambda}{2}}}{\sqrt{1-t}}(a^\lambda+a^{-\lambda}) & {\rm otherwise}.
\end{array}\rt.
\ea
It is conjectured in \cite{Song-TQFT} that the general wavefunction for $I_{2,2n}$ is given by the following scaling rule, 
\ba
f^{I_{2,2m}}_\lambda(a;t)=\lt\{\begin{array}{cc}
\sqrt{\frac{1+t}{1-t}} & \lambda=0,\\
\frac{t^{\frac{\lambda}{2}(m+1)}}{\sqrt{1-t}}(a^\lambda+a^{-\lambda}) & {\rm otherwise}.
\end{array}\rt.
\ea
One can check the expressions of wavefunction for irregular punctures by comparing the HL index with the 3d Coulomb branch index of the 3d mirror theory developed in \cite{3dmirror}. For example, with 
\ba
C^{HL\ -1}_\lambda=\frac{\tilde{P}^{HL}_\lambda(t^{\rho}|t)}{\prod_{i=1}^r(1-t^{d_i})},
\ea
and especially for $n=2$, 
\ba
C^{HL\ -1}_\emptyset=\frac{1}{\sqrt{1-t^2}},\\
C^{HL\ -1}_\lambda=\frac{t^{-\frac{\lambda}{2}}}{\sqrt{1-t}},
\ea
we can check that 
\ba
I^{(A_1,A_{-2})}_{HL}(t)=\sum_\lambda C^{HL\ -1}_\lambda f^{I_{2,-1}}_\lambda(t)=0,\\
I^{(A_1,A_{2n-2})}_{HL}(t)=\sum_\lambda C^{HL\ -1}_\lambda f^{I_{2,2n-1}}_\lambda(t)=1,
\ea
which indicate $(A_1,A_{-2})$ theory is trivial and $(A_1,A_{2n-2})$ theories for $n\geq 1$ have no Higgs branch.

For $n=3$, the wavefunction of $I_{3,-2}$ is given by 
\ba
f^{I_{3,-2}}_\lambda(t)=\frac{1}{3!}\oint\frac{{\rm d}x_1}{2\pi ix_1}\frac{{\rm d}x_2}{2\pi ix_2}\Delta(x_1,x_2)I^{HL}_{vec}(z_1,z_2;t)f^{HL}_\lambda(z_1,z_2)\nn\\
=\frac{1}{3!}\oint\frac{{\rm d}x_1}{2\pi ix_1}\frac{{\rm d}x_2}{2\pi ix_2}\Delta(x_1,x_2)\tilde{P}^{HL}_\lambda(x_1,x_2),
\ea
where the Haar measure $\Delta(x_i)$ is given by 
\ba
\Delta(x_i)=\prod_{i\neq j}(1-x_i/x_j),
\ea
and we see that the integral above extracts out the term $\chi_\emptyset^{su(3)}(z_1,z_2)$ in the HL polynomial $P^{HL}_\lambda(x_1,x_2)$. Using the expressions of the HL polynomials expanded in terms of SU(3) characters, (\ref{HL-schur-s})-(\ref{HL-schur-e}), we obtain for $|\lambda|\leq 4$, 
\ba
f^{I_{3,-2}}_{\emptyset}(t)=\sqrt{(1-t^2)(1-t^3)},\\
f^{I_{3,-2}}_{(2,1)}(t)=-t(1-t^2),\\
f^{I_{3,-2}}_{(3,0)}(t)=t^2\sqrt{(1-t)(1-t^2)},
\ea
and zero for other configurations. It is not difficult to see that the wavefunction vanishes unless the representation of $\lambda$ contains the Cartan part, i.e. the highest weight of the representation should be of the form 
\ba
\vec{w}=(3k,3\ell)\ {\rm or}\ (3k-2,3\ell-2),
\ea
for some non-negative integers $k$ and $\ell$. We note that this observation is also true in the Schur limit. By further using 
\ba
C^{HL\ -1}_\lambda=\frac{\tilde{P}^{HL}_\lambda(t,1,t^{-1}|t)}{(1-t^2)(1-t^3)},
\ea
we have for example 
\ba
C^{HL\ -1}_\emptyset=\frac{1}{\sqrt{(1-t^2)(1-t^3)}},\\
C^{HL\ -1}_{(2,1)}=\frac{t^{-2}}{1-t},\\
C^{HL\ -1}_{(3,0)}=\frac{t^{-3}}{1-t^2}\sqrt{1+t}.
\ea
To recover the trivial Higgs branch of $(A_2,A_{m-1})$ theories for $m$ that is not a multiple of 3, it is natural to conjecture that 
\ba
f^{I_{3,m}}_{\lambda}(t)=\sqrt{(1-t^2)(1-t^3)}\delta_{\lambda,\emptyset}.
\ea

One consistency check can be performed to compute the HL index for pure SU(3) YM, as 
\ba
I^{pure SU(3)}_{HL}=\sum_\lambda f_\lambda^{I_{3,-2}}(t)f_\lambda^{I_{3,-2}}(t)=(1-t^2)(1-t^3)+t^2(1-t^2)^2+2t^4(1-t)(1-t^2)+\dots\nn\\
=1-t^3+{\cal O}(t^5),
\ea
which agrees with the direct computation 
\ba
I^{pure SU(3)}_{HL}=\frac{1}{3!}\oint\frac{{\rm d}z_1}{2\pi iz_1}\frac{{\rm d}z_2}{2\pi iz_2}\Delta(z_1,z_2)I^{HL}_{vec}(z_1,z_2;t)\nn\\
=\frac{(1-t)^2}{3!}\oint\frac{{\rm d}z_1}{2\pi iz_1}\frac{{\rm d}z_2}{2\pi iz_2}\prod_{\alpha\in\Delta}(1-z^\alpha)(1-tz^\alpha)\nn\\
=1-t^3-t^5+t^8.
\ea

\section{Wavefunction in Macdonald Limit}\label{s:Mac-wave}

The wavefunction for (full) regular puncture in the Macdonald limit is given by 
\ba
f_\lambda(z)=\frac{\tilde{P}_\lambda(z)}{(t;q)^r\prod_{\alpha\in\Delta}(tz^\alpha;q)}.
\ea
The wavefunction satisfy again the normalization condition 
\ba
\oint\prod_i \frac{{\rm d}z_i}{2\pi iz_i}\prod_{\alpha\in\Delta}(1-z^\alpha)I_{vec}(z)f_{\lambda}(z)f_{\mu}(z^{-1})=\delta_{\lambda\mu},
\ea
where 
\ba
I_{vec}(z)=(q;q)^r(t;q)^r\prod_{\alpha\in\Delta}(qz^\alpha;q)(tz^\alpha;q).
\ea
We can see that the Macdonald polynomial used here is normalized s.t. 
\ba
\frac{1}{n!}\frac{(q;q)^r}{(t;q)^r}\oint\prod_i \frac{{\rm d}z_i}{2\pi iz_i}\prod_{\alpha\in\Delta}\frac{(z^\alpha;q)}{(tz^\alpha;q)}\tilde{P}_\lambda(z)\tilde{P}_\mu(z^{-1})=\delta_{\lambda\mu},
\ea
and it can be expressed in terms of the Macdonald polynomial $\bar{P}_\lambda(x|q,t)$ we present in Appendix \ref{a:Macdonald} as 
\ba
\tilde{P}_\lambda(z)=\frac{(t;q)^{\frac{r}{2}}}{(q;q)^{\frac{r}{2}}}\bar{P}_\lambda(x|q,t).
\ea

We can work out the wavefunction for the irregular puncture $I_{2,-1}$ through the integral 
\ba
f^{I_{2,-1}}_\lambda(q,t)&=&\frac{1}{2}\oint\frac{{\rm d}\zeta}{2\pi i\zeta}(1-\zeta^2)(1-\zeta^{-2})I_{vec}(\zeta;q,t)f_\lambda(\zeta)\nn\\
&=&(q;q)\frac{1}{2}\oint\frac{{\rm d}\zeta}{2\pi i\zeta}(\zeta^2;q)(\zeta^{-2};q)\tilde{P}_\lambda(\zeta)\nn\\
&=&\frac{1}{2}\frac{(t;q)^{\frac{1}{2}}}{(q;q)^{\frac{1}{2}}}\oint\frac{{\rm d}\zeta}{2\pi i\zeta}(q;q)(\zeta^2;q)(\zeta^{-2};q)\bar{P}_\lambda(\zeta).
\ea
When we decompose the Macdonald polynomial in terms of SU($n$) characters, we can evaluate the above integral (for a general rank) as discussed in \cite{Song-TQFT}, 
\ba
&&I_\lambda:=\frac{(q;q)^r}{n!}\oint\prod_i \frac{{\rm d}z_i}{2\pi iz_i}\prod_{\alpha\in\Delta}(z^\alpha;q)\chi_\lambda(z)\nn\\
&&=\frac{1}{|\cW|(q;q)^{\frac{1}{2}(n-1)(n-2)}}\sum_{n_\alpha\in\mathbb{Z}}(-1)^{\sum_{\alpha\in\Delta_+} n_\alpha}q^{\sum_{\alpha\in\Delta_+}\frac{1}{2}n_\alpha(n_\alpha+1)}\sum_{\sigma\in\cW}\epsilon(\sigma)\delta_{\sigma\cdot w=n_\alpha \alpha},
\ea
where $\sigma\cdot w=\sigma(w+\rho)-\rho$ is the shifted Weyl reflection. For $n=2$, the above integral can be evaluated to 
\ba
I_\lambda=\lt\{\begin{array}{cc}
(-1)^{\frac{\lambda}{2}}q^{\frac{1}{2}\frac{\lambda}{2}(\frac{\lambda}{2}+1)} & \lambda:\ {\rm even},\\
0 & \lambda:\ {\rm odd}.\\
\end{array}\rt.
\ea
The wavefunction of $I_{2,-1}$ can then be calculated to 
\ba
f^{I_{2,-1}}_0(q,t)=\lt(\frac{(t^2;q)}{(tq;q)}\rt)^{\frac{1}{2}},\quad f^{I_{2,-1}}_2(q,t)=-t\lt(\frac{(1-t)(1-q^2)(t^2q^2;q)}{(1-q)(1-tq)(tq^3;q)}\rt)^{\frac{1}{2}},\\
f^{I_{2,-1}}_4(q,t)=t^2q\lt(\frac{(t;q)_2(q^3;q)_2(t^2q^4;q)}{(q;q)_2(tq^2;q)_2(tq^5;q)}\rt)^{\frac{1}{2}},\quad \dots,
\ea
and in general we have \cite{Song-TQFT} 
\ba
f^{I_{2,-1}}_\lambda(q,t)=(-1)^{\frac{\lambda}{2}}t^{\frac{\lambda}{2}}q^{\frac{1}{2}\frac{\lambda}{2}(\frac{\lambda}{2}-1)}\lt(\frac{(t;q)_{\frac{\lambda}{2}}(q^{\frac{\lambda}{2}+1};q)_{\frac{\lambda}{2}}(t^2q^{\lambda};q)}{(q;q)_{\frac{\lambda}{2}}(tq^{\frac{\lambda}{2}};q)_{\frac{\lambda}{2}}(tq^{\lambda+1};q)}\rt)^{\frac{1}{2}},
\ea
for $\lambda$ even, and zero for $\lambda$ odd. In the limit $q\rightarrow 0$, i.e. the HL limit, only the wavefunction for $\lambda=0$ and $\lambda=2$ survive. However, as we know that the wavefunction for $I_{2,2i+1}$ with $i\geq 0$ has to vanish unless $\lambda=0$, the scaling rule in the Macdonald limit for the wavefunction can be speculated (together with the wavefunction in the Schur limit given in \cite{Song-TQFT}) to 
\ba
f^{I_{2,2i+1}}_\lambda(q,t)=(-1)^{\frac{\lambda}{2}}q^{\frac{\lambda}{2}(\frac{\lambda}{2}+1)(i+\frac{3}{2})}(t/q)^{\frac{\lambda}{2}(i+2)}\lt(\frac{(t;q)_{\frac{\lambda}{2}}(q^{\frac{\lambda}{2}+1};q)_{\frac{\lambda}{2}}(t^2q^{\lambda};q)}{(q;q)_{\frac{\lambda}{2}}(tq^{\frac{\lambda}{2}};q)_{\frac{\lambda}{2}}(tq^{\lambda+1};q)}\rt)^{\frac{1}{2}}.\label{wave-rank-one}
\ea

For $n=3$, one can argue that only when the highest weight of the representation of $\lambda$ of the form 
\ba
\vec{w}=(3k,3\ell)\ {\rm or}\ (3k-2,3\ell-2),\label{contributing-w}
\ea
we can obtain a non-zero wavefunction for $I_{3,-2}$. We have 
\ba
I_\lambda(q)=\lt\{\begin{array}{cc}
q^{k(k+1)+\ell(\ell+1)+k\ell} & w_1=3k,\ w_2=3\ell,\\
-q^{k^2+\ell^2-1+(k-1)(\ell-1)} & w_1=3k-2,\ w_2=3\ell-2,\\
0 & {\rm otherwise},\\
\end{array}\rt.
\ea
which leads to 
\ba
&&f^{I_{3,-2}}_\emptyset(q,t)=\lt(\frac{(1-t^2)(t^3;q)}{(tq;q)}\rt)^{\frac{1}{2}},\\
&&f^{I_{3,-2}}_{(2,1)}(q,t)=-\frac{t(1-t)(1+t+qt)}{1-q^2t}\lt(\frac{(1-t^2q^2)(t^3q^2;q)}{(tq^3;q)}\rt)^{\frac{1}{2}},\\
&&f^{I_{3,-2}}_{(3,0)}(q,t)=f^{I_{3,-2}}_{(3,3)}(q,t)=t^2\frac{(1-q^2)(1-q^3)}{(1-tq)(1-tq^2)}\lt(\frac{(1-t)(t^2;q)_4(t^3q^3;q)}{(q;q)_3(tq^3;q)}\rt)^{\frac{1}{2}}.
\ea

The structure constant 
\ba
C^{-1}_\lambda=\frac{\tilde{P}_\lambda(t^\rho)}{\prod_{i=1}^r (t^{d_i};q)},
\ea
for $n=2$ is given by 
\ba
C^{-1}_\lambda=t^{-\frac{\lambda}{2}}\lt(\frac{1}{(t;q)_\lambda(q;q)_\lambda(tq^{\lambda+1};q)(t^2q^\lambda;q)}\rt)^{\frac{1}{2}},
\ea
and for $n=3$ can be read as 
\ba
&&C^{-1}_\emptyset=\frac{1}{(t^2q;q)}\lt(\frac{1}{(1-t^2)(tq;q)(t^3;q)}\rt)^{\frac{1}{2}},\\
&&C^{-1}_{(2,1)}=\frac{t^{-2}(1+t)}{(1-q)(1-tq^2)(t^2;q)}\lt(\frac{1-t^2q^2}{(t^3q^2;q)(tq^3;q)}\rt)^{\frac{1}{2}},\\
&&C^{-1}_{(3,0)}=\frac{t^{-3}}{(t^2;q)}\lt(\frac{(t^2;q)_4}{(q;q)_3(tq;q)_2(t;q)(t^3q^3;q)}\rt)^{\frac{1}{2}}.
\ea

The most naive conjecture about the wavefunction of $I_{3,2}$ is then given by
\ba
&&f^{I_{3,2}}_\emptyset(q,t)=\lt(\frac{(1-t^2)(t^3;q)}{(tq;q)}\rt)^{\frac{1}{2}},\label{wavefunction-A21-dual}\\
&&f^{I_{3,2}}_{(2,1)}(q,t)=-\frac{t^4q(1-t)(1+q+qt)}{1-q^2t}\lt(\frac{(1-t^2q^2)(t^3q^2;q)}{(tq^3;q)}\rt)^{\frac{1}{2}},\label{wavefunction-A22-32}\label{wavefunction-A22-dual}\\
&&f^{I_{3,2}}_{(3,0)}(q,t)=f^{I_{3,2}}_{(3,3)}(q,t)=t^6q^{4}\lt[\frac{(1-q^2)(1-q^3)}{(1-tq)(1-tq^2)}\lt(\frac{(1-t)(t^2;q)_4(t^3q^3;q)}{(q;q)_3(tq^3;q)}\rt)^{\frac{1}{2}}+{\cal O}(q^4)\rt],\nn\\\label{wavefunction-A23-dual}
\ea
where we required only wavefunctions for representations satisfying (\ref{contributing-w}) can be non-vanishing, and the above expressions reproduce the $(A_1,A_2)$ index, as we will soon check. 

As will be discussed in section \ref{s:dis}, we conjecture that the first two wavefunctions for $I_{3,m}$ singularity with ${\rm gcd}(3,m)=0$ are of the form 
\ba
&&f^{I_{3,m}}_\emptyset(q,t)=\lt(\frac{(1-t^2)(t^3;q)}{(tq;q)}\rt)^{\frac{1}{2}},\label{wavefunction-A21}\\
&&f^{I_{3,m}}_{(2,1)}(q,t)=-\frac{t^{2j+k+2}q^{1+j}(1-t)(1+(t/q)^{\delta_{k,1}}q+qt)}{1-q^2t}\lt(\frac{(1-t^2q^2)(t^3q^2;q)}{(tq^3;q)}\rt)^{\frac{1}{2}},\label{wavefunction-A22}
\ea
where we set $m=3j+k$ with $k=1,2$. Note that as pointed out in \cite{ALS}, the pattern of the Pltethystic exponential expression of the Macdonald indices for $(A_2,A_{m-1})$ theories differs in the case $k=1$ and $k=2$, so it is natural to have the wavefunction beyond $\lambda=\emptyset$ to depend on $\delta_{k,1}$. 

In the same spirit, for $n=4$, we have 
\ba
f^{I_{4,m}}_\emptyset(q,t)=\lt(\frac{(1-t^2)(1-t^3)(t^4;q)}{(tq;q)}\rt)^{\frac{1}{2}},
\ea
with ${\rm gcd}(4,m)=1$, and it is expected that for large enough $m$, 
\ba
I^{(A_3,A_{m-1})}=1+Tq^2+(T+T^2)q^3+(T+2T^2+T^3)q^4+(T+2T^2+2T^3)q^5+{\cal O}(q^6).
\ea
We will give a closed form formula for the Macdonald index of $(A_{n-1},A_{m-1})$ in the large $m$ limit based on the wavefunction of $I_{n,-n+1}$ for $\lambda=\emptyset$ in section \ref{s:dis}. 

On the other hand, one can also work out the wavefunction for $I_{2,0}$ in the Macdonald limit as we did in the HL limit, and then extrapolate it to $I_{2,2m}$ by matching with the expressions in Schur and HL limit. This has already been done in \cite{BN}, and the result is 
\ba
f^{I_{2,2m}}_\lambda(x;q,t)=\frac{t^{\frac{(m+1)\lambda}{2}}q^{\frac{(m+1)\lambda^2}{4}}}{(t;q)}\sum_{i=0}^\lambda \frac{(t;q)_i(t;q)_{\lambda-i}}{(q;q)_i(q;q)_{\lambda-i}}q^{-(m+1)(\frac{\lambda}{2}-i)^2}x^{2i-\lambda}\lt(\frac{(q;q)_\lambda(t^2q^\lambda;q)}{(t;q)_\lambda(tq^{\lambda+1};q)}\rt)^{\frac{1}{2}}.
\ea
We will in particular be interested in the $(A_1,A_1)$ theory, which is nothing but a free hypermultiplet theory, in this article. 

Let us check the level-rank duality between $(A_1,A_2)$ theory and $(A_2,A_1)$ theory to conclude this section. One can compute the index of $(A_1,A_2)$ theory from two different points of view. As the $(A_1,A_2)$ theory built from 6d theory with $G=A_1$, we have 
\ba
&&I^{(A_1,A_2)}=\sum_\lambda C^{-1}_\lambda f_\lambda^{I_{2,3}}=\sum_{m=0}^\infty (-1)^m\frac{q^{\frac{5}{2}m^2-\frac{1}{2}m}t^{2m}}{(tq^{2m+1};q)(tq^m;q)_m(q;q)_m}\nn\\
&&=1+Tq^2+Tq^3+Tq^4+Tq^5+(T+T^2)q^6+(T+T^2)q^7+(T+2T^2)q^8+(T+2T^2)q^9\nn\\
&&+(T+3T^2)q^{10}+{\cal O}(q^{11}),\label{Mac-25}
\ea
where we put $T:=t/q$ in the above expansion. We have on the dual side with $G=A_2$, 
\ba
&&I^{(A_2,A_1)}=\sum_\lambda C^{-1}_\lambda f_\lambda^{I_{3,2}}=1+Tq^2+Tq^3+Tq^4+Tq^5+(T+T^2)q^6\nn\\
&&+(T+T^2)q^7+(T+2T^2)q^8+(T+2T^2)q^9+(T+3T^2)q^{10}+{\cal O}(q^{11}),
\ea
which reproduces the $(A_1,A_2)$ Macdonald index up to order $q^{10}$. 

\section{Higgsing Prescription}\label{s:Higgsing}

Let us review the Higgsing prescription first formulated in \cite{Higgsing} and then put forward in \cite{NSZ} to fit into the class S picture, which is a useful method to generate surface operators in 4d $\cN=2$ gauge theories with no Lagrangian description. 

Given a target theory we want to insert surface operators into, we first prepare a UV theory, which can be obtained by adding one more regular puncture to the TQFT construction of the original theory. In the current context, we want to investigate Argyres-Douglas theories constructed from the compactification of 6d $\cN=(2,0)$ theory on a sphere with an irregular puncture of type $I_{n,m}$, so the UV theory has a dual TQFT picture with a regular puncture in addition to the irregular puncture $I_{n,m}$ on a sphere. The index of the UV theory in the Macdonald limit can be computed as 
\ba
I^{UV}(q,t;z)=\sum_\lambda f_\lambda(z)f^{I_{n,m}}_\lambda(q,t).
\ea
There is an additional SU($n$) flavor symmetry in this UV theory (associated to the regular puncture), and one is allowed to Higgs this flavor symmetry with the moment map operators attached to it. As the Schur operator $(\sigma^\mu_{+\dot{+}}\partial_\mu)^{s_i}{\cal O}_{\alpha_i}$ associated to the momentum map of simple root $\alpha_i$ has charge $(\Delta,j_1,j_2,R,r)=(2+s_i,\frac{s_i}{2},\frac{s_i}{2},1,0)$, its contribution to the Macdonald index is given by 
\ba
\frac{1}{1-tq^{s_i}z^{\alpha_i}},\label{mm-contr}
\ea
and to Higgs this operator, i.e. giving a position-dependent VEV to the flavor symmetry, we need to take the limit $z\rightarrow z_\ast$, where $z_\ast$ satisfies 
\ba
tq^{s_i}z^{\alpha_i}_\ast=1,
\ea
and get rid of some appropriate (divergent) decoupled d.o.f.'s when flow back to the original (IR) SCFT. One therefore obtain the original theory with a surface operator sitting at the origin of the chiral algebra plane, which can be labeled by a set of non-negative integer numbers $\{s_i\}$. Its Macdonald index can be computed via 
\ba
I^{\mathbb{S}^{\{s_i\}}}_{\cal T}(q,t)=F^{\{s_i\}}_{\cal T}(q,t)\lim_{z\rightarrow z_\ast}\lt(I^{UV}(q,t;z)\prod_{1\leq i\leq j\leq n-1}(1-\frac{t}{q}\prod_{\ell=i}^jq^{s_\ell+1}z^{\alpha_\ell})\rt),
\ea
where the strip-off factor $F^{\{s_i\}}_{\cal T}(q,t)$ was presumably proposed in \cite{Higgsing} to be worked out from the decoupling of a number of U(1) free hypermultiplets. As the Macdonald index of a free U(1) hypermultiplet is given by\footnote{We note that in fact the position of poles in (\ref{mm-contr}) does not match exactly with that in the free U(1) hypermultiplet index.} 
\ba
\frac{1}{(t^{\frac{1}{2}}z^{\pm 1};q)},
\ea
the strip-off factor for the case of $n=2$ is expected to be proportional to 
\ba
(q;q)(q;q)_s(tq^s;q),\label{naive-guess-strip}
\ea
where the coefficient in $F^{\{s_i\}}_{\cal T}(q,t)$ is of the form $\alpha t^\beta q^\gamma$ with some $\alpha,\beta,\gamma\in\mathbb{R}$ s.t. the refined character is normalized properly. We note that this is the same prescription as (A.10) in \cite{CGS-chiral} in the Macdonald limit. 

In the Schur limit, it was shown in \cite{NSZ} that $\{s\}$ gives the most natural label of modules in the dual chiral algebra. In the series of $(A_{n-1},A_{m-1})$ with ${\rm gcd}(n,m)=1$, $\{s\}$ coincides with the label of minimal model modules $\{s_i\}_{i=1}^n$ as will be reviewed in section \ref{s:Song}. 

One can compare our results for $(A_1,A_1)$ theory with the results shown in \cite{CGS-chiral}. We also refer to \cite{Dedushenko:2019yiw} for a relevant analysis. The UV index for the $(A_1,A_1)$ theory is given by 
\ba
I^{UV}(z,x;q,t)=\sum_\lambda \frac{t^\lambda q^{\frac{1}{2}\lambda^2} (t^2q^\lambda;q)P_\lambda(z)}{(t;q)^2(tq^{\lambda+1};q)(tz^2;q)(tz^{-2};q)}\sum_{i=0}^\lambda \frac{(t;q)_i(t;q)_{\lambda-i}}{(q;q)_i(q;q)_{\lambda-i}}q^{-2(\frac{\lambda}{2}-i)^2}x^{2i-\lambda}.
\ea
The Macdonald index of the $(A_1,A_1)$ theory is then found to be 
\ba
I_{(A_1,A_1)}(x;q,t)=1+T^{\frac{1}{2}}(x+x^{-1})q^{\frac{1}{2}}+T(x^2+1+x^{-2})q+T^{\frac{1}{2}}(x+x^{-1})(Tx^2+1+Tx^{-2})q^{\frac{3}{2}}\nn\\
+T\lt((x+x^{-1})^2+T(x^4+x^2+1+x^{-2}+x^{-4})\rt)q^2+{\cal O}(q^{\frac{5}{2}}),
\ea
which matches with the expansion of 
\ba
I_{(A_1,A_1)}(x;q,t)=\frac{1}{(t^{\frac{1}{2}}x;q)(t^{\frac{1}{2}}x^{-1};q)},
\ea
over $q$, i.e the index of the free hypermultiplet. The Macdonald index for $s=1$ with the strip-off factor (\ref{naive-guess-strip}) can be computed in the same way to 
\ba
&&I^{\mathbb{S}^1}_{(A_1,A_1)}(x;q,t)=(1+\sqrt{T}(x+x^{-1})+T(x^2+x^{-2})+T^{\frac{3}{2}}(x^3+x^{-3})+\dots)\nn\\
&&+\lt(\sqrt{T}(x+x^{-1})+T(x+x^{-1})^2+T^{\frac{3}{2}}(x^3+x+x^{-1}+x^{-3})+T^2(x^4+x^2+x^{-2}+x^{-4})+\dots\rt)q\nn\\
&&+\lt(\sqrt{T}(x+x^{-1})+T(2x^2+3+2x^{-2})+T^{\frac{3}{2}}(2x^3+3x+3x^{-1}+2x^{-3})+\dots\rt)q^2+{\cal O}(q^3),
\ea
which is exactly the expansion of 
\ba
I^{\mathbb{S}^1}_{(A_1,A_1)}(x;q,t)=\frac{1-t/q}{(t^{\frac{1}{2}}q^{-\frac{1}{2}}x;q)(t^{\frac{1}{2}}q^{-\frac{1}{2}}x^{-1};q)},\label{mac-11-1}
\ea
presented in \cite{CGS-chiral}, and also reproduces the correct answer of the Schur index 
\ba
I^{\mathbb{S}^1}_{(A_1,A_1)}(x;q,q)=\frac{\delta(x)}{(qx;q)(qx^{-1};q)},
\ea
where the $\delta$-function is defined as $\delta(x)=\sum_{i\in\mathbb{Z}}x^i$. 

We will save the computation of $s=2$ in $(A_1,A_1)$ to the discussion part, section \ref{s:dis}, together with the chiral algebra interpretation of the above results. 

\section{Macdonald Index with Surface Operator}\label{s:surface}

Let us start to compute the Macdonald index with the known expressions of wavefunctions and the conjecture (\ref{wavefunction-A21})-(\ref{wavefunction-A22}) and the strip-off factor (\ref{naive-guess-strip}) in this section. We will compare the results with the refined character in the section \ref{s:dis}. 

The Macdonald index for a $(A_1,A_{2i})$ theory with no surface operator inserted is given by 
\ba
&&I^{(A_1,A_{2i})}=\sum_\lambda C^{-1}_\lambda f_\lambda^{I_{2,2i+1}}=\sum_{m=0}^\infty (-1)^m\frac{q^{(i+\frac{3}{2})m^2-\frac{1}{2}m}t^{(i+1)m}}{(tq^{2m+1};q)(tq^m;q)_m(q;q)_m}.
\ea
Let us list the explicit expressions for several examples beyond the $(A_1,A_2)$ theory here. 
\ba
I^{(A_1,A_{4})}&=&1+Tq^2+Tq^3+(T+T^2)q^4+(T+T^2)q^5+(T+2T^2)q^6+(T+2T^2)q^7\nn\\
&&+(T+3T^2+T^3)q^8+(T+3T^2+2T^3)q^9+(T+4T^2+3T^3)q^{10}+{\cal O}(q^{11}),\label{index-14}\\
I^{(A_1,A_{6})}&=&1+Tq^2+Tq^3+(T+T^2)q^4+(T+T^2)q^5+(T+2T^2+T^3)q^6+(T+2T^2+T^3)q^7\nn\\
&&+(T+3T^2+2T^3)q^8+(T+3T^2+3T^3)q^9+(T+4T^2+4T^3+T^4)q^{10}+{\cal O}(q^{11}),\label{index-16}\\
I^{(A_1,A_{8})}&=&1+Tq^2+Tq^3+(T+T^2)q^4+(T+T^2)q^5+(T+2T^2+T^3)q^6+(T+2T^2+T^3)q^7\nn\\
&&+(T+3T^2+2T^3+T^4)q^8+(T+3T^2+3T^3+T^4)q^9+(T+4T^2+4T^3+2T^4)q^{10}+{\cal O}(q^{11}).\nn\\\label{index-18}
\ea

For $n=3$, let us deal with three examples, $(A_2,A_3)$, $(A_2,A_4)$ and $(A_2,A_6)$ theory. We note that up to ${\cal O}(q^{10})$, their indices can be computed with only the wavefunction of (\ref{wavefunction-A21}) and (\ref{wavefunction-A22}). 
\ba
I^{(A_2,A_{3})}&=&1+Tq^2+(T+T^2)q^3+(T+2T^2)q^4+(T+2T^2)q^5+(T+3T^2+2T^3)q^6\nn\\
&&+(T+3T^2+3T^3)q^7+(T+4T^2+5T^3+T^4)q^8+(T+4T^2+7T^3+2T^4)q^9+{\cal O}(q^{10}).\nn\\\label{index-23}
\ea
\ba
I^{(A_2,A_{4})}&=&1+Tq^2+(T+T^2)q^3+(T+2T^2)q^4+(T+2T^2+T^3)q^5+(T+3T^2+3T^3)q^6\nn\\
&&+(T+3T^2+4T^3)q^7+(T+4T^2+6T^3+3T^4)q^8+(T+4T^2+8T^3+5T^4)q^9+{\cal O}(q^{10}).\nn\\\label{index-24}\\
I^{(A_2,A_{6})}&=&1+Tq^2+(T+T^2)q^3+(T+2T^2)q^4+(T+2T^2+T^3)q^5+(T+3T^2+3T^3+T^4)q^6\nn\\
&&+(T+3T^2+4T^3+2T^4)q^7+(T+4T^2+6T^3+5T^4)q^8\nn\\
&&+(T+4T^2+8T^3+7T^4+2T^5)q^9+{\cal O}(q^{10}).\label{index-26}
\ea
We note that they all agree with the results obtained in \cite{ALS} from the VOA approach to $W_3$-algebra plus selection rules. 

Now we go back to the case of $n=2$ and consider the insertion of surface operator. We need to put the flavor fugacity parameter to $z_\ast=t^{\frac{1}{2}}q^{\frac{s}{2}}$ to introduce a surface operator labeled by $s$. In the $(A_1,A_2)$ theory, one expects the $s=1$ surface operator to reproduce the $h=-\frac{1}{5}$ module of the Lee-Yang singularity. 
\ba
I^{\mathbb{S}^1}_{(A_1,A_2)}(q,t)&=&\sum_\lambda \frac{\tilde{P}_\lambda(t^{\frac{1}{2}}q^{\frac{1}{2}})}{(1-t)(t^2q;q)}f^{I_{2,3}}_\lambda(q,t)=\sum_{m=0}^\infty (-1)^mq^{\frac{5}{2}m^2-\frac{1}{2}m}t^{3m}\frac{P_{2m}(t^{\frac{1}{2}}q^{\frac{1}{2}})}{(1-t)(t^2q;q)}\frac{(t;q)_m(t^2q^{2m};q)}{(q;q)_m(tq^{2m+1};q)}\nn\\
&=&1+Tq+Tq^2+Tq^3+(T+T^2)q^4+(T+T^2)q^5+(T+2T^2)q^6+(T+2T^2)q^7\nn\\
&&+(T+3T^2)q^8+(T+3T^2+T^3)q^9+(T+4T^2+T^3)q^{10}+{\cal O}(q^{11}),\label{Mac-25-1}
\ea
which matches with the refined character computed in (\ref{ref-ch-251}). 

In general, the Macdonald index for $(A_1,A_{2i})$ theory with $s=1$ surface operator inserted can be computed to 
\ba
I^{\mathbb{S}^1}_{(A_1,A_{2i})}(q,t)=\sum_{m=0}^\infty (-1)^mq^{(i+\frac{3}{2})m^2-\frac{1}{2}m}t^{m(i+2)}\frac{P_{2m}(t^{\frac{1}{2}}q^{\frac{1}{2}})}{(1-t)(t^2q;q)}\frac{(t;q)_m(t^2q^{2m};q)}{(q;q)_m(tq^{2m+1};q)},
\ea
and in particular, 
\ba
I^{\mathbb{S}^1}_{(A_1,A_4)}(q,t)&=&1+Tq+Tq^2+(T+T^2)q^3+(T+2T^2)q^4+(T+2T^2)q^5+(T+3T^2+T^3)q^6\nn\\
&&+(T+3T^2+2T^3)q^7+(T+4T^2+3T^3)q^8+(T+4T^2+5T^3)q^9\nn\\
&&+(T+5T^2+6T^3+T^4)q^{10}+{\cal O}(q^{11}),\label{Mac-14-1}\\
I^{\mathbb{S}^1}_{(A_1,A_6)}(q,t)&=&1+Tq+Tq^2+(T+T^2)q^3+(T+2T^2)q^4+(T+2T^2+T^3)q^5+(T+3T^2+2T^3)q^6\nn\\
&&+(T+3T^2+3T^3)q^7+(T+4T^2+4T^3+T^4)q^8+(T+4T^2+6T^3+2T^3)q^9\nn\\
&&+(T+5T^2+7T^3+4T^4)q^{10}+{\cal O}(q^{11}).\label{Mac-16-1}
\ea

The general Macdonald index for the series $(A_1,A_{2i})$ with $s=2$ surface operator inserted and the naive strip-off factor (\ref{naive-guess-strip}) is given by 
\ba
I^{\mathbb{S}^2}_{(A_1,A_{2i})}(q,t)=\sum_{m=0}^\infty (-1)^mq^{(i+\frac{3}{2})m^2-\frac{1}{2}m}t^{m(i+2)}\frac{P_{2m}(t^{\frac{1}{2}}q)}{(1-t)(1-tq)(t^2q^2;q)}\frac{(t;q)_m(t^2q^{2m};q)}{(q;q)_m(tq^{2m+1};q)}.
\ea
More explicitly, 
\ba
I^{\mathbb{S}^2}_{(A_1,A_{2})}(q,t)&=&1+Tq+(2T-T^2)q^2+Tq^3+(T+3T^2-2T^3)q^4+(T+2T^2-T^3)q^5\nn\\
&&+(T+3T^2+T^3-2T^4)q^6+(T+3T^2-T^4)q^7+{\cal O}(q^8),\\
I^{\mathbb{S}^2}_{(A_1,A_{4})}(q,t)&=&1+Tq+2Tq^2+(T+T^2)q^3+(T+4T^2-2T^3)q^4+(T+3T^2)q^5\nn\\
&&+(T+4T^2+4T^3-3T^4)q^6+(T+4T^2+4T^3-2T^4)q^7+{\cal O}(q^8),\\
I^{\mathbb{S}^2}_{(A_1,A_{6})}(q,t)&=&1+Tq+2Tq^2+(T+T^2)q^3+(T+4T^2-T^3)q^4+(T+3T^2+T^3)q^5\nn\\
&&+(T+4T^2+5T^3-3T^4)q^6+(T+4T^2+5T^3-T^4)q^7+{\cal O}(q^8).
\ea
For $s=4$, we have 
\ba
I^{\mathbb{S}^4}_{(A_1,A_{2})}(q,t)=(1-T^2)+(T-T^2)q+(2T-T^2-T^3)q^2+(2T-2T^3)q^3\nn\\
+(2T+3T^2-5T^3)q^4+{\cal O}(q^5),\\
I^{\mathbb{S}^4}_{(A_1,A_{4})}(q,t)=1+Tq+(2T-T^3)q^2+(2T+T^2-T^3)q^3\nn\\
+(2T+4T^2-2T^3-T^4)q^4+{\cal O}(q^5).
\ea
We note the above indices can never be interpreted as a refined character of the minimal models. 

Let us conclude this section by giving a primitive discussion on the computation of Macdonald index with surface operators inserted for $n=3$. The first guess for the strip-off factor is\footnote{We remark, however, that in general only two of three $(\sigma^\mu\partial_\mu)^{s_i}\cO_{\alpha_i}$ for $\alpha_i\in\Delta_+$ become massless in the Macdonald limit, while all of them truly decouple in the Schur limit.} 
\ba
(q;q)^2(q;q)_{s_1}(tq^{s_1};q)(q;q)_{s_2}(tq^{s_2};q)(t^{-1}q^2;q)(t;q)_{s_1+s_2+1}(t^2q^{s_1+s_2};q).\label{strip-off-rank2}
\ea
The Macdonald index can then be computed via 
\ba
I^{\mathbb{S}^{s_1,s_2}}_{(A_2,A_{m-1})}(q,t)= \sum_\lambda \frac{(t^2q^{s_1+s_2};q)\tilde{P}_\lambda(tq^{\frac{2}{3}s_1+\frac{1}{3}s_2},tq^{\frac{1}{3}s_1+\frac{2}{3}s_2})}{(t;q)_{s_1}(t;q)_{s_2}(t^2q^{s_1};q)(t^2q^{s_1};q)(t^3q^{s_1+s_2};q)}f^{I_{3,m}}_\lambda(q,t).
\ea
For sufficiently large $m$, e.g. $m\geq 7$, we have for $(s_1=1,s_2=0)$, 
\ba
I^{\mathbb{S}^{1,0}}_{(A_2,A_{m-1})}(q,t)=1+Tq+(T+T^2)q^2+(T+2T^2)q^3+(T+3T^2+2T^3)q^4\nn\\
+(T+3T^2+4T^3+T^4)q^5+{\cal O}(q^6).
\ea

\section{Song's Prescription for Chiral Algebra}\label{s:Song}

Local operators contributing to the superconformal index in the Schur limit coincide with those contributing in a looser limit, the Macdonald limit. This motivates one to guess that the Macdonald index gives certain kind of refined character to the chiral algebra introduced in \cite{ChiralSym}. This idea, which was first explored in \cite{Song}, worked surprisingly well for the rank one AD theories at the level of vaccum character. Let us review how Song's prescription works for Virasoro minimal models, which are dual to $(A_1,A_{2m})$ theories, with concrete examples in the first half of this section, and we provide an alternative approach to the refined character based on the POSET description of minimal models in the latter half. 

The central charge for a general $(p=n,q)$ minimal model is given by \cite{Fateev-Lukyanov}
\ba
c=(n-1)\lt(1-\frac{(p-q)^2}{pq}n(n+1)\rt).
\ea

In the simplest case of Lee-Yang edge singularity, $(p,q)=(2,5)$, $c=-\frac{22}{5}$. One can compute the vacuum character by identifying the spectrum of null states, where there is always a primary null state $L_{-1}\ket{0}$. The next primary null state appears at level 4, given by 
\ba
(L_{-4}-\frac{5}{3}L^2_{-2})\ket{0}.
\ea
Its descendant states at level 5 and level 6 can be computed to 
\ba
(L_{-5}-\frac{5}{2}L_{-3}L_{-2})\ket{0},\quad (L_{-6}+\frac{1}{2}L_{-4}L_{-2}-\frac{5}{6}L_{-2}^3)\ket{0},\quad (L_{-6}-\frac{5}{4}L_{-4}L_{-2}-\frac{5}{8}L^2_{-3})\ket{0}.
\ea
One can thus convert states with more numbers of Virasoro generators to the linear combination of those with less by using the above expressions of null states to obtain the following non-singular spectrum, 
\ba
&&{\rm level\ 2}:\quad L_{-2}\ket{0},\nn\\
&&{\rm level\ 3}:\quad L_{-3}\ket{0},\nn\\
&&{\rm level\ 4}:\quad L_{-4}\ket{0},\nn\\
&&{\rm level\ 5}:\quad L_{-5}\ket{0},\nn\\
&&{\rm level\ 6}:\quad L_{-6}\ket{0},\quad L_{-4}L_{-2}\ket{0},\\
&&\dots\nn
\ea
The refined character that corresponds to the Macdonald index of a Virasoro module like this is given by 
\ba
\chi(q,T)={\rm tr}\lt((-1)^FT^\# q^h\rt),
\ea
where $\#$ stands for the number of Virasoro generators needed to construct a state. The refined character for the vacuum module of the Lee-Yang singualrity is thus given by 
\ba
\chi^{(2,5)}_{s=0}(q,T)=1+Tq^2+Tq^3+Tq^4+Tq^5+(T+T^2)q^6+{\cal O}(q^7),
\ea
which indeed agrees with the Macdonald index we computed in (\ref{Mac-25}) for $(A_1,A_2)$ theory. 


One can work out the refined character for the $h=-\frac{1}{5}$ module of the Lee-Yang singularity in the same way. The first two primary null states can be found to 
\ba
(L_{-2}-\frac{5}{2}L_{-1}^2)\ket{-1/5},\quad (L_{-3}-\frac{25}{12}L_{-1}^3)\ket{-1/5},
\ea
which restrict the spectrum together with their descendants to 
\ba
&&{\rm level\ 1}:\quad L_{-1}\ket{-1/5},\nn\\
&&{\rm level\ 2}:\quad L_{-2}\ket{-1/5},\nn\\
&&{\rm level\ 3}:\quad L_{-3}\ket{-1/5},\nn\\
&&{\rm level\ 4}:\quad L_{-4}\ket{-1/5},\quad L_{-2}^2\ket{-1/5},\nn\\
&&{\rm level\ 5}:\quad L_{-5}\ket{-1/5},\quad L_{-4}L_{-1}\ket{-1/5},\nn\\
&&{\rm level\ 6}:\quad L_{-6}\ket{-1/5},\quad L_{-4}L_{-2}\ket{-1/5},\quad L_{-3}^2\ket{-1/5},\\
&&\dots\nn
\ea
The refined character for this module reads  
\ba
\chi^{(2,5)}_{s=1}(q,T)=1+Tq+Tq^2+Tq^3+(T+T^2)q^4+(T+T^2)q^5+(T+2T^2)q^6+{\cal O}(q^7).\label{ref-ch-251}
\ea

It is also intriguing to work out the refined character for the vacuum module of $(p,q)=(3,5)$ theory, which is a level-rank dual description of the Lee-Yang singularity. At $c=-\frac{22}{5}$, in addition to $L_{-1}\ket{0,0}$, $W_{-1}\ket{0,0}$, $W_{-2}\ket{0,0}$, we have $W_{-3}\ket{0,0}$ as a new primary null state, which kills all the states excited by the $W$-generators, and thus we obtain the same spectrum. We remark that even for $h=-\frac{1}{5}$, $W_{-1}\ket{0,0}$, $W_{-2}\ket{0,0}$ and $W_{-3}\ket{0,0}$ are null again, so the dual $\cW_3$ description has indeed the same spectrum as the Virasoro Lee-Yang singularity model. 

The above method can immediately be generalized to all modules of all Virasoro minimal models. It is, however, not easy in general to work out the spectrum and deal with all the null states carefully up to high orders. Here we provide an alternative approach to the refined character: the partially ordered set (POSET) \cite{Stanley:1986:EC:21786} approach. The POSET method was developed in \cite{FNMZ,Foda:2015bsa} to compute the character of $\cW$-algebraic models with singular vectors. 

As a well-known fact, the highest weight module of a $(p,q)$ $\cW_n$ minimal model is labeled by two sets of $n$-positive-component vectors, $(n_i)$ and $(n_i')$, satisfying 
\ba
\sum_{i=1}^nn_i=q,\quad \sum_{i=1}^nn_i'=p.
\ea
As shown in \cite{NSZ}, when $p=n$, $n'_i$'s become trivial, and $n_i$ is related to the label of surface operator through 
\ba
n_i=1+s_i.
\ea
The conformal weight for the highest weight state of this module is given by 
\ba
h(s)=\frac{1}{2q}\lt(\sum_{i=1}^{n-1}i(n-i)(s_i^2-(q-n)s_i)+2\sum_{i<j}^{n-1}i(n-j)s_is_j\rt).
\ea
We further introduce a set of integer numbers, which can be found from the equations, 
\ba
a_i=\sum_{j=1}^{i-1}(j-i)s_j-\sum_{j=i}^n(n-j+i)s_j,\quad i=1,\dots,n,
\ea
and an infinite set of integer numbers generated from $a_i$'s by 
\ba
\{a_i+q-pk\},\quad k\in\mathbb{Z}_{>0}.
\ea
Let us denote the above infinite set as 
\ba
{\cal X}:=\{a_i+q-pk\mid i=1,\dots,n,k\in\mathbb{Z}_{>0}\},
\ea
then we can consider a partition 
\ba
\pi=\{\pi(m)\}_{m\in{\cal X}},
\ea
obeying a partial ordering, 
\ba
\pi(m)\geq \pi(m-p),\quad \pi(m)\geq \pi(m-(q-p)),
\ea
for $^\forall m\in{\cal X}$. The Hilbert space of all physical states in the corresponding module of $\cW_n$-algebra multiplied with a $\mathfrak{u}(1)$ Heisenberg algebra can be identified to the space of all such partitions $\pi$. The character of the $\cW_n$ minimal model module can thus be computed by 
\ba
\chi=(q;q)\sum_\pi q^{|\pi|},
\ea
where we multiplied a factor of $(q;q)$ to decouple the $\mathfrak{u}(1)$ d.o.f. 

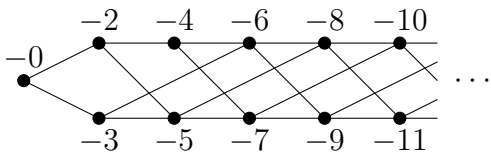
\begin{figure}[H]
\centering
\begin{tikzpicture}[scale=1]
\foreach \q in {0,1}
\foreach \p in {0,1,...,4}
\pgfmathtruncatemacro{\i}{2*\p+\q+2}
\coordinate (\i)at (\p,-\q);
\coordinate (0)at (-1,-.5);
\foreach \r in {3,5,...,11}
\draw[fill] (\r) circle[radius=.08] node[below] {$-\r$};
\foreach \r in {0,2,...,10}
\draw[fill] (\r) circle[radius=.08] node[above] {$-\r$};
\foreach \s in {0,2,3,...,8}
\pgfmathtruncatemacro{\j}{\s+3}
\draw (\s)--(\j);
\foreach \s in {0,2,3,...,9}
\pgfmathtruncatemacro{\j}{\s+2}
\draw (\s)--(\j);
\draw (10)--([xshift=.5cm]10);
\draw (11)--([xshift=.5cm]11);
\draw (10)--([xshift=.5cm,yshift=-.5cm]10);
\draw (11)--([xshift=.5cm,yshift=.25cm]11);
\draw (9)--([xshift=1.5cm,yshift=.75cm]9);
\node at (5, -.5) {$\cdots$};
\end{tikzpicture}
\caption{The POSET diagram for $h=0$ in Lee-Yang model.}
\label{Poset-0}
\end{figure}

Let us look at the simplest example, the vacuum module of the $(p,q)=(2,5)$ Lee-Yang singularity. We have $(s_1,s_2)=(0,2)$ for this module and therefore 
\ba
{\cal X}=\mathbb{Z}_{\leq 0}\backslash\{-1\}.
\ea
One can draw a diagram shown in Figure \ref{Poset-0} to help understand the situation, where when two numbers $j,k\in{\cal X}$ are connected by a line, it means the partition to the number on the left, say $j$, is larger than or equal to that of the number on the right, say $k$, i.e. 
\ba
\pi(j)\geq \pi(k).
\ea
We can for example fix a canonical ordering, e.g. $(0,-2,-3,-4,-5,-6,-7,\dots)$, so that 
\ba
\pi(0)\geq\pi(-2)\geq \pi(-3)\geq \pi(-4)\geq\pi(-5)\geq\pi(-6)\geq \pi(-7)\geq \dots
\ea
whose contribution to the $\mathfrak{u}(1)\times \cW_2$ character is 
\ba
\chi(0,-2,-3,-4,-5,-6,-7,\dots)=\frac{1}{(q;q)},
\ea
which is exactly the character of a normal partition.

All the other allowed partitions that do not involve double counting will then have one or more $>$ (without equality) in the corresponding characteristic ordering. For example, 
\ba
\pi(0)\geq\pi(-3)> \pi(-2)\geq \pi(-4)\geq\pi(-5)\geq\pi(-6)\geq \pi(-7)\geq \dots\label{1st-label}
\ea
is the next simplest ordering one can consider for the Lee-Yang vacuum module. The contribution from this ordering, $(0,-3,-2,-4,-5,-6,-7,\dots)$, is 
\ba
\chi(0,-3,-2,-4,-5,-6,-7,\dots)=\frac{q^2}{(q;q)}.
\ea
More generally, even for other models and higher-rank algebras, the contribution from the ordering $(m_1,m_2,m_3,\dots)$ with $\pi(m_i)>\pi(m_{i+1})$ for a set of positive integers $\{i\}$ is given by 
\ba
\chi(m_1,m_2,m_3,\dots)=\frac{\prod_{\{i\}} q^i}{(q;q)}.
\ea
Back to the case of the vacuum module of the Lee-Yang singularity, we have 
\ba
\chi^{(2,5)}_{s=0}(q)=(q;q)\sum_{(n_i):\ {\rm allowed}}\chi(n_1,n_2,n_3,\dots)=\sum_{k=0}^\infty\sum_{\substack{t_1,t_2,\dots,t_k\\t_{i+1}-t_i\geq 2\\t_1\geq 2}}q^{\sum_{i=1}^kt_i}=\sum_{k=0}^\infty \frac{q^{k^2+k}}{(q;q)_k}=\frac{1}{(q^2;q^5)(q^3;q^5)},\nn\\
\ea
where in the last line we used the second Rogers-Ramanujan identity. 

Our prescription for the {\bf refined character} when $n=2$ is given as follows. We first pair all neighboring numbers in the canonical ordering, for instance, in the example above, we have pairs $(0,-2)$, $(-2,-3)$, $(-3,-4)$, $(-4,-5)$, \dots We refine the contribution of each ordering by counting the number of such pairs above, $(m_i,m_{i+1})$, satisfying 
\ba
\pi(m_{i+1})>\pi(m_i),
\ea
while each number $m_i$ can be only paired once and we maximize the number of such pairing in our counting. By denoting this number as $\#(\mathfrak{o})$ for an allowed ordering $\mathfrak{o}$, the refined character of this ordering $\mathfrak{o}$ is given by 
\ba
\chi(\mathfrak{o};q,T)=T^{\#(\mathfrak{o})}\frac{\prod_{\{i\}_\mathfrak{o}} q^i}{(q;q)}.
\ea
The full refined character of the module under consideration then reads 
\ba
\chi(q,T)=(q;q)\sum_{\mathfrak{o}:\ {\rm allowed}}\chi(\mathfrak{o};q,T).
\ea

In the case of Lee-Yang singularity, $\#(\mathfrak{o})$ matches with the number of $>$ in the ordering $\mathfrak{o}$, and the vacuum refined character can be found from 
\ba
\chi^{(2,5)}_{s=0}(q,T)&=&\sum_{k=0}^\infty T^k\sum_{\substack{t_1,t_2,\dots,t_k\\t_{i+1}-t_i\geq 2\\t_1\geq 2}}q^{\sum_{i=1}^kt_i}\nn\\
&=&1+Tq^2+Tq^3+Tq^4+Tq^5+(T+T^2)q^6+(T+T^2)q^7+(T+2T^2)q^8\nn\\
&&+(T+2T^2)q^9+(T+3T^2)q^{10}+(T+3T^2)q^{11}+{\cal O}(q^{12}).\label{refined-Lee-Yang}
\ea
For example, $T^3$-term with the lowest order comes from the ordering 
\ba
(0,-3_>,-2,-5_>,-4,-7_>,-6,-8,-9,-10,-11,\dots),
\ea
where we added $>$ signs to indicate the locations equality no longer holds in the corresponding partition, and it contributes $T^3q^{12}$ to the refined character. The above result perfectly matches with the result we obtained from the Macdonald index and Song's prescription. 

For the $h=-\frac{1}{5}$ module of the Lee-Yang singularity model, ${\cal X}=\mathbb{Z}_{<0}$, 
and one can compute the refined character for this model in the same way. This time, as the ordering $(-2,-1,-3,-4,-5,-6,-7,\dots)$ is allowed, we have a non-trivial contribution $Tq$ at level 1. The full refined character reads 
\ba
\chi^{(2,5)}_{s=1}(q,T)&=&1+Tq+Tq^2+Tq^3+(T+T^2)q^4+(T+T^2)q^5+(T+2T^2)q^6+(T+2T^2)q^7\nn\\
&&+(T+3T^2)q^8+(T+3T^2+T^3)q^9+(T+4T^2+T^3)q^{10}+{\cal O}(q^{11}),
\ea
which again agrees perfectly with the Macdonald index we computed for $(A_1,A_2)$ theory with $s=1$ surface operator inserted, (\ref{Mac-25-1}), and the refined character in Song's prescription.

The infinite set $\mathcal{X}$ can be computed for any $(2,2+m)$ model to be ${\cal X}=\{-2i\}_{i\in\mathbb{Z}_{\geq 0}}\cup\{-m-2i\}_{i\in\mathbb{Z}_{\geq 0}}$. We can then compute the vacuum refined character of $(2,7)$, $(2,9)$ and $(2,11)$ minimal models following the prescription described above. For example, in the $(2,7)$ model, we obtain the following list of the contributions from the first several orderings, 
\ba
&&(0,-2,-4,-5,-6,-7,-8,-9,-10,-11,-12,\dots)\mapsto 1,\nn\\
&&(0,-5_>,-2,-4,-6,-7,-8,-9,-10,-11,-12,\dots)\mapsto Tq^2,\nn\\
&&(0,-2,-5_>,-4,-6,-7,-8,-9,-10,-11,-12,\dots)\mapsto Tq^3,\nn\\
&&(0,-2,-4,-6_>,-5,-7,-8,-9,-10,-11,-12,\dots)\mapsto Tq^4,\nn\\
&&(0,-2,-5,-7_>,-4,-6,-8,-9,-10,-11,-12,\dots)\mapsto T^2q^4.
\ea 
We obtain the refined characters for $(2,7)$, $(2,9)$ and $(2,11)$ minimal models in this way as 
\ba
\chi^{(2,7)}_{s=0}(q,T)&=&1+Tq^2+Tq^3+(T+T^2)q^4+(T+T^2)q^5+(T+2T^2)q^6+(T+2T^2)q^7\nn\\
&&+(T+3T^2+T^3)q^8+{\cal O}(q^{9}),\\
\chi^{(2,9)}_{s=0}(q,T)&=&1+Tq^2+Tq^3+(T+T^2)q^4+(T+T^2)q^5+(T+2T^2+T^3)q^6+(T+2T^2+T^3)q^7\nn\\
&&+(T+3T^2+2T^3)q^8+{\cal O}(q^{9}),\\
\chi^{(2,11)}_{s=0}(q,T)&=&1+Tq^2+Tq^3+(T+T^2)q^4+(T+T^2)q^5+(T+2T^2+T^3)q^6+(T+2T^2+T^3)q^7\nn\\
&&+(T+3T^2+2T^3+T^4)q^8+{\cal O}(q^{9}).
\ea
The above results completely agree respectively with the Macdonald index for $(A_1,A_4)$, (\ref{index-14}), the index for $(A_1,A_6)$, (\ref{index-16}) and the index $(A_1,A_8)$, (\ref{index-18}). 

${\cal X}$ is found to be ${\cal X}=\{-1-2i\}_{i\in\mathbb{Z}_{\geq 0}}\cup\{1-m-2i\}_{i\in\mathbb{Z}_{\geq 0}}$ for the $s=1$ module of $(2,2+m)$ model. The refined character of this module in $(2,7)$ and $(2,9)$ models are respectively given by 
\ba
\chi^{(2,7)}_{s=1}(q,T)&=&1+Tq+Tq^2+(T+T^2)q^3+(T+2T^2)q^4+(T+2T^2)q^5+(T+3T^2+T^3)q^6\nn\\
&&+(T+3T^2+2T^3)q^7+(T+4T^2+3T^3)q^8+(T+4T^2+5T^3)q^9+{\cal O}(q^{10}),\\
\chi^{(2,9)}_{s=1}(q,T)&=&1+Tq+Tq^2+(T+T^2)q^3+(T+2T^2)q^4+(T+2T^2+T^3)q^5+(T+3T^2+2T^3)q^6\nn\\
&&+(T+3T^2+3T^3)q^7+(T+4T^2+4T^3+T^4)q^8+(T+4T^2+6T^3+2T^4)q^9+{\cal O}(q^{10}),\nn\\
\ea
which match with the results from the Macdonald index computation (\ref{Mac-14-1}) and (\ref{Mac-16-1}).

${\cal X}=\{-2-2i\}_{i\in\mathbb{Z}_{\geq 0}}\cup\{2-m-2i\}_{i\in\mathbb{Z}_{\geq 0}}$ for the $s=2$ module of $(2,2+m)$ model. The refined characters of this module in $(2,7)$ and $(2,9)$ models are given by 
\ba
\chi^{(2,7)}_{s=2}(q,T)&=&1+Tq+(T+T^2)q^2+(T+T^2)q^3+(T+2T^2)q^4+(T+2T^2+T^3)q^5+(T+3T^2+2T^3)q^6\nn\\
&&+(T+3T^2+3T^3)q^7+(T+4T^2+4T^3+T^4)q^8+(T+4T^2+6T^3+T^4)q^9+{\cal O}(q^{10}),\label{ref-27-2}\\
\chi^{(2,9)}_{s=2}(q,T)&=&1+Tq+(T+T^2)q^2+(T+T^2)q^3+(T+2T^2+T^3)q^4+(T+2T^2+2T^3)q^5\nn\\
&&+(T+3T^2+3T^3)q^6+(T+3T^2+4T^3+T^4)q^7+(T+4T^2+5T^3+3T^4)q^8+{\cal O}(q^9)\label{ref-29-2}.
\ea

The naive extension of this prescription to higher rank case does not give the correct refined character in Song's sense. We will discuss a possible improvement of our method for higher rank $\cW$-algebras in the next section. 

\section{Comparison}\label{s:dis}

As we have already seen in section \ref{s:surface}, the naive computation of Macdonald index with surface operators inserted does not always give the refined character of the corresponding module of the chiral algebra. There are several possible ways to explain this tension: 

\begin{itemize}
\item Song's prescription for the Macdonald index to be a refined character only miraculously holds for some special and simple modules. 

\item The strip-off factor (\ref{naive-guess-strip}) needs to be modified. 

\item The refinement rule for the refined character differs from module to module. 

\item The expressions for wavefunctions need to be revised. 

\end{itemize}

\subsection{$(A_1,A_1)$ theory revisited}

Let us check firstly whether Song's prescription always only works for $s=0,1$. To do so, we go back to the simplest Lagrangian case of $(A_1,A_1)$ theory. The dual chiral algebra is generated by the $\beta\gamma$-system. In the vacuum module generated from the highest weight $\ket{0}$, which is annihilated by $\beta_{i}\ket{0}=0$, $\gamma_{i}\ket{0}=0$, for $i\in\frac{1}{2}\mathbb{Z}_{>0}$, one can have arbitrary numbers of $\beta_{-i}$ and $\gamma_{-i}$ excited from the vacuum $\ket{0}$, i.e. the character is given by 
\ba
\chi^{\beta\gamma}_{s=0}(x;q)=\frac{1}{(q^{\frac{1}{2}}x;q)(q^{\frac{1}{2}}x^{-1};q)},
\ea
where $x$ is the fugacity parameter for the U(1) current\footnote{The convention here is that $[\beta_i,\gamma_j]=\delta_{n+m,0}$.} $J(z)=-:\gamma\beta:(z)$, under which $\beta$ and $\gamma$ are respectively charged $\pm 1$. Song's prescription for the $\beta\gamma$ system is to further count the number of generators $\#(\beta\gamma)$, $\beta_{-i}$ and $\gamma_{-i}$, contained in each state, and to refine the character with $T^{\frac{1}{2}\#(\beta\gamma)}$ \cite{Song}. The refined character given by this prescription reads 
\ba
\chi^{\beta\gamma}_{s=0}(x;q,T)=\frac{1}{((qT)^{\frac{1}{2}}x;q)((qT)^{\frac{1}{2}}x^{-1};q)}=\frac{1}{(t^{\frac{1}{2}}x;q)(t^{\frac{1}{2}}x^{-1};q)},
\ea
and it matches exactly with the Macdonald index of a free hypermultiplet. The module corresponding to $s=1$ can be obtained from a spectral flow using the U(1) current $J$ from the vacuum module. Note that now $\beta$ and $\gamma$ currents have integer expansion modes, there are two possible highest weight states to be considered: i) $\ket{+}_0$ annihilated by $\gamma_0$ and all positive modes; ii) $\ket{-}_0$ annihilated by $\beta_0$ and all positive modes. The whole module can be generated from $\beta_0^n\ket{+}$ and $\gamma_0^n\ket{-}$ for $^\forall n\in\mathbb{Z}_{\geq 0}$ with excitations of $\beta_{-i}$ and $\gamma_{-i}$ for $i\in\mathbb{Z}_{>0}$. The character can thus be computed as 
\ba
\chi^{\beta\gamma}_{s=1}(x;q)=\lt(\frac{1}{1-x}+\frac{x^{-1}}{1-x^{-1}}\rt)\frac{1}{(qx;q)(qx^{-1};q)}=\frac{\delta(x)}{(qx;q)(qx^{-1};q)}=I^{\mathbb{S}^1}_{(A_1,A_1)}(x;q,q).
\ea
The refined character under Song's prescription is then given by 
\ba
\chi^{\beta\gamma}_{s=1}(x;q,T)=\lt(\frac{1}{1-T^{\frac{1}{2}}x}+\frac{T^{\frac{1}{2}}x^{-1}}{1-T^{\frac{1}{2}}x^{-1}}\rt)\frac{1}{(T^{\frac{1}{2}}qx;q)(T^{\frac{1}{2}}qx^{-1};q)}=\frac{1-T}{(T^{\frac{1}{2}}x;q)(T^{\frac{1}{2}}x^{-1};q)},
\ea
which agrees with the Macdonald index $I^{\mathbb{S}^1}_{(A_1,A_1)}(x;q,t)$, (\ref{mac-11-1}). For $s=2$, the Schur index carried out with our Higgsing method reads 
\ba
I^{\mathbb{S}^2}_{(A_1,A_1)}(x;q,q)&=\dots&+\lt((x^2+x^{-2})+2(x^4+x^{-4})+5(x^6+x^{-6})+10(x^8+x^{-8})+\dots\rt)q^{-1}\nn\\
&&+\lt((x+x^{-1})+2(x^3+x^{-3})+5(x^5+x^{-5})+10(x^7+x^{-7})+\dots\rt)q^{-\frac{1}{2}}\nn\\
&&+\lt(1+2(x^2+x^{-2})+5(x^4+x^{-4})+10(x^6+x^{-6})+\dots\rt)\nn\\
&&+\lt(2(x+x^{-1})+5(x^3+x^{-3})+10(x^5+x^{-5})+20(x^7+x^{-7})+\dots\rt)q^{\frac{1}{2}}\nn\\
&&+\lt(3+5(x^2+x^{-2})+10(x^4+x^{-4})+20(x^6+x^{-6})+36(x^8+x^{-8})+\dots\rt)q\nn\\
&&+\lt(5(x+x^{-1})+10(x^3+x^{-3})+20(x^5+x^{-5})+36(x^7+x^{-7})+\dots\rt)q^{\frac{3}{2}}\nn\\
&&+\lt(7+10(x^2+x^{-2})+20(x^4+x^{-4})+36(x^6+x^{-6})+\dots\rt)q^{\frac{3}{2}}\nn\\
&&+{\cal O}(q^{\frac{5}{2}}),
\ea
where now the character is not even bounded from below. It is in fact a mixture of three irreducible modules. One is the vacuum module generated from $\ket{0}$, while the negative powers of $q$ come from $\ket{\pm}_{\frac{1}{2}}$ satisfying 
\ba
\beta_i\ket{+}_{\frac{1}{2}}=0,\quad \gamma_j\ket{+}_{\frac{1}{2}}=0,\\
\beta_j\ket{-}_{\frac{1}{2}}=0,\quad \gamma_i\ket{-}_{\frac{1}{2}}=0,
\ea 
for $i=\frac{3}{2},\frac{5}{2},\frac{7}{2},\dots$ and $j=-\frac{1}{2},\frac{1}{2},\frac{3}{2},\frac{5}{2},\dots$ The characters of the modules associated to $\ket{\pm}_{\frac{1}{2}}$ are respectively given by 
\ba
\chi^{\beta\gamma}_{\pm\frac{1}{2}}(x;q)=\frac{1}{1-x^\pm q^{-\frac{1}{2}}}\frac{1}{(x^\pm q^{\frac{1}{2}};q)(x^{\mp}q^{\frac{3}{2}};q)}.
\ea
As an expansion over $q$, we have 
\ba
\chi^{\beta\gamma}_{+\frac{1}{2}}(x;q)&=\dots&+\lt(x^2+2x^4+5x^6+10x^8+20x^{10}+36x^{12}+\dots\rt)q^{-1}\nn\\
&&+\lt(x+2x^3+5x^5+10x^7+20x^{9}+36x^{11}+\dots\rt)q^{-\frac{1}{2}}\nn\\
&&+\lt(1+2x^2+5x^4+10x^6+20x^{8}+36x^{10}+\dots\rt)\nn\\
&&+\lt(2x+5x^3+10x^5+20x^7+36x^9+\dots\rt)q^{\frac{1}{2}}\nn\\
&&+\lt(1+5x^2+10x^4+20x^6+36x^8+\dots\rt)q^1\nn\\
&&+\lt(x^{-1}+4x+10x^3+20x^5+36x^7+\dots\rt)q^{\frac{3}{2}}\nn\\
&&+\lt(3+9x^2+20x^4+36x^6+\dots\rt)q^2+{\cal O}(q^{\frac{5}{2}}),
\ea
and $\chi^{\beta\gamma}_{-\frac{1}{2}}(q)$ is given by replacing $x$ with $x^{-1}$ in the above expression. One can check that 
\ba
I^{\mathbb{S}^2}_{(A_1,A_1)}(x;q,q)=\chi^{\beta\gamma}_{s=0}(q)+xq^{-\frac{1}{2}}\chi^{\beta\gamma}_{+\frac{1}{2}}(q)+x^{-1}q^{-\frac{1}{2}}\chi^{\beta\gamma}_{-\frac{1}{2}}(q),
\ea
up to high orders. We can also try to be sloppy (analytically continue) to rewrite 
\ba
\chi^{\beta\gamma}_{\pm\frac{1}{2}}(x;q)=-\frac{x^{\mp}q^{\frac{1}{2}}}{(x q^{\frac{1}{2}};q)(x^{-1}q^{\frac{1}{2}};q)},
\ea
and then we arrive at 
\ba
I^{\mathbb{S}^2}_{(A_1,A_1)}(x;q,q)=-\frac{1}{(x q^{\frac{1}{2}};q)(x^{-1}q^{\frac{1}{2}};q)}=-\chi^{\beta\gamma}_{s=0}(q),
\ea
which matches with the result in \cite{CGS-chiral}. We remark that the ``periodicity" of the Schur index in the case of $(A_1,A_1)$ theory only appears after an analytic continuation in our approach, and does not appear at the level of computation with surface operator insertion directly. The Macdonald index computed from the strip-off factor (\ref{naive-guess-strip}) is given by 
\ba
I^{\mathbb{S}^2}_{(A_1,A_1)}(x;q,t)&=\dots&+\lt(T(x^2+x^{-2})+(T^2+T^3)(x^4+x^{-4})+(2T^3+2T^4+T^5)(x^6+x^{-6})+\dots\rt)q^{-1}\nn\\
&&+\lt(\sqrt{T}(x+x^{-1})+(T^{\frac{3}{2}}+T^{\frac{5}{2}})(x^3+x^{-3})+\dots\rt)q^{-\frac{1}{2}}\nn\\
&&+\lt(1+(T+T^2)(x^2+x^{-2})+(2T^2+2T^3+T^4)(x^4+x^{-4})+\dots\rt)\nn\\
&&+\lt((T^{\frac{1}{2}}+T^{\frac{3}{2}})(x+x^{-1})+(2T^{\frac{3}{2}}+2T^{\frac{5}{2}}+T^{\frac{7}{2}})(x^3+x^{-3})+\dots\rt)q^{\frac{1}{2}}\nn\\
&&+\lt(3T+(2T+2T^2+T^3)(x^2+x^{-2})+\dots\rt)q+{\cal O}(q^{\frac{3}{2}}).
\ea
Following Song's prescription, we obtain 
\ba
\sqrt{T}\chi^{\beta\gamma}_{+\frac{1}{2}}(x;q,T)&=&\frac{T^{\frac{1}{2}}}{1-xT^{\frac{1}{2}}q^{-\frac{1}{2}}}\frac{1}{(x T^{\frac{1}{2}}q^{\frac{1}{2}};q)(x^{-1} T^{\frac{1}{2}}q^{\frac{3}{2}};q)}\nn\\
&=&\dots+\lt(T^{\frac{3}{2}}x^2+(T^{\frac{5}{2}}+T^{\frac{7}{2}})x^4+\dots\rt)q^{-1}\nn\\
&&+\lt(Tx+(T^2+T^3)x^2+(2T^3+2T^4+T^5)x^4+\dots\rt)q^{-\frac{1}{2}}\nn\\
&&+\lt(\sqrt{T}+(T^{\frac{3}{2}}+T^{\frac{5}{2}})x^2+(2T^{\frac{5}{2}}+2T^{\frac{7}{2}}+T^{\frac{9}{2}})x^4+\dots\rt)\nn\\
&&+\lt((T+T^2)x+(2T^2+2T^3+T^4)x^3+\dots\rt)q^{\frac{1}{2}}\nn\\
&&+\lt(T^{\frac{3}{2}}+(2T^{\frac{3}{2}}+2T^{\frac{5}{2}}+T^{\frac{7}{2}})x^2+\dots\rt)q\nn\\
&&+\lt(Tx^{-1}+(T+2T^2+T^3)x^2+\dots\rt)q^{\frac{3}{2}}+{\cal O}(q^2),
\ea
($\chi^{\beta\gamma}_{-\frac{1}{2}}(x;q,T)$ again given by $x\rightarrow x^{-1}$) and together with 
\ba
\chi^{\beta\gamma}_{s=0}(x;q,T)=1+T^{\frac{1}{2}}(x+x^{-1})q^{\frac{1}{2}}+T(x^2+1+x^{-2})q+T^{\frac{1}{2}}(x+x^{-1})(Tx^2+1+Tx^{-2})q^{\frac{3}{2}}\nn\\
+T\lt((x+x^{-1})^2+T(x^4+x^2+x^{-2}x^{-4})\rt)q^2+{\cal O}(q^{\frac{5}{2}}),
\ea
it appears that 
\ba
I^{\mathbb{S}^2}_{(A_1,A_1)}(x;q,t)=\chi^{\beta\gamma}_{s=0}(x;q,T)+\sqrt{T}xq^{-\frac{1}{2}}\chi^{\beta\gamma}_{+\frac{1}{2}}(x;q,T)+\sqrt{T}x^{-1}q^{-\frac{1}{2}}\chi^{\beta\gamma}_{-\frac{1}{2}}(x;q,T),
\ea
which is a perfect realization of Macdonald index as Song's refined character. 

\subsection{$(A_{n-1},A_{m-1})$ theories with ${\rm gcd}(n,m)=1$ at large $m$}

Now we turn to investigate the vacuum character of $G=A_2$ (rank 2) case. As we discussed before, for large enough $m$, (e.g. $m\geq 10$), the first several contributions (e.g. up to order ${\cal O}(q^{10})$) are completely determined by the wavefunction 
\ba
f^{I_{3,m}}_\emptyset(q,t)=\lt(\frac{(1-t^2)(t^3;q)}{(tq;q)}\rt)^{\frac{1}{2}},
\ea
and therefore 
\ba
I^{(A_2,A_{m-1})}&=&I^{(A_2,A_{6})}+{\cal O}(q^{8})=\frac{1}{(T^2q^3;q)(Tq^2;q)}+{\cal O}(q^{10})\nn\\
&=&1+Tq^2+(T+T^2)q^3+(T+2T^2)q^4+(T+2T^2+T^3)q^5+(T+3T^2+3T^3+T^4)q^6\nn\\
&&+(T+3T^2+4T^3+2T^4)q^7+(T+4T^2+6T^3+5T^4+T^5)q^8\nn\\
&&+(T+4T^2+8T^3+7T^4+3T^5+T^6)q^9+{\cal O}(q^{10}).
\ea
Of course, in the $m\rightarrow \infty$ limit, $f^{I_{3,m}}_\emptyset(q,t)$ completely determined the Macdonald index. The universal spectrum of $(p=3,q\geq 10)$ is in fact given by\footnote{We quote the paper of \cite{W3rep} for more details on such computations.} 
\ba
{\rm level\ 0}:\ &&\ket{0},\nn\\
{\rm level\ 2}:\ &&L_{-2}\ket{0},\nn\\
{\rm level\ 3}:\ &&L_{-3}\ket{0},\quad W_{-3}\ket{0},\nn\\
{\rm level\ 4}:\ &&L_{-4}\ket{0},\quad L_{-2}L_{-2}\ket{0},\quad W_{-4}\ket{0},\nn\\
{\rm level\ 5}:\ &&L_{-5}\ket{0},\quad L_{-3}L_{-2}\ket{0},\quad W_{-5}\ket{0},\quad W_{-3}L_{-2}\ket{0},\nn\\
{\rm level\ 6}:\ &&L_{-6}\ket{0},\quad L_{-4}L_{-2}\ket{0},\quad L_{-3}L_{-3}\ket{0}\quad W_{-6}\ket{0},\nn\\
&&L_{-2}L_{-2}L_{-2}\ket{0},\quad W_{-4}L_{-2}\ket{0},\quad W_{-3}L_{-3}\ket{0},\quad W_{-3}^2\ket{0},\nn\\
{\rm level\ 7}:\ &&L_{-7}\ket{0},\quad L_{-5}L_{-2}\ket{0},\quad L_{-4}L_{-3}\ket{0}\quad W_{-7}\ket{0},\nn\\
&&L_{-3}L_{-2}L_{-2}\ket{0},\quad W_{-5}L_{-2}\ket{0},\quad W_{-4}L_{-3}\ket{0},\quad W_{-3}L_{-4}\ket{0},\nn\\
&&W_{-4}W_{-3}\ket{0},\quad W_{-3}L_{-2}^2\ket{0},\nn\\
{\rm level\ 8}:\ &&L_{-8}\ket{0},\quad L_{-5}L_{-3}\ket{0},\quad L_{-4}L_{-4}\ket{0},\quad L_{-6}L_{-2}\ket{0}, \quad W_{-8}\ket{0},\nn\\
&&L_{-4}L_{-2}L_{-2}\ket{0},\quad L_{-3}L_{-3}L_{-2}\ket{0},\quad W_{-6}L_{-2}\ket{0},\quad W_{-5}L_{-3}\ket{0},\quad W_{-3}L_{-5}\ket{0},\quad W_{-4}L_{-4}\ket{0},\nn\\
&&W_{-5}W_{-3}\ket{0},\quad W_{-4}W_{-4}\ket{0},\quad W_{-4}L_{-2}^2\ket{0},\quad W_{-3}L_{-3}L_{-2}\ket{0},\quad L_{-2}^4\ket{0},\nn\\
&&W_{-3}^2L_{-2}\ket{0},\nn
\ea
\ba
{\rm level\ 9}:\ &&L_{-9}\ket{0},\quad L_{-6}L_{-3}\ket{0},\quad L_{-5}L_{-4}\ket{0},\quad L_{-7}L_{-2}\ket{0}, \quad W_{-9}\ket{0},\nn\\
&&L_{-5}L_{-2}L_{-2}\ket{0},\quad L_{-4}L_{-3}L_{-2}\ket{0},\quad W_{-7}L_{-2}\ket{0},\quad W_{-6}L_{-3}\ket{0},\quad W_{-4}L_{-5}\ket{0},\quad W_{-5}L_{-4}\ket{0},\nn\\
&&W_{-3}L_{-6}\ket{0},\quad L^3_{-3}\ket{0},\quad W_{-6}W_{-3}\ket{0},\quad W_{-5}W_{-4}\ket{0},\quad W_{-5}L_{-2}^2\ket{0},\quad W_{-4}L_{-3}L_{-2}\ket{0},\nn\\
&&W_{-3}L_{-4}L_{-2}\ket{0},\quad W_{-3}L_{-3}L_{-3}\ket{0},\quad L_{-3}L^3_{-2}\ket{0},\nn\\
&&W_{-3}^2L_{-3}\ket{0},\quad W_{-4}W_{-3}L_{-2}\ket{0},\quad W_{-3}L_{-2}^3\ket{0},\quad W_{-3}^3\ket{0},
\ea
which exactly reproduces the Macdonald index if we count the number of $W$-generators with weight $T^2$ and Virasoro operators with weight $T$. This can be directly seen from the exact formula in the large $m$ limit that 
\ba
\lim_{m\rightarrow \infty}I^{(A_2,A_{m-1})}=\frac{1}{(Tq^2;q)(T^2q^3;q)},\label{large-m-2}
\ea
where the spectrum is excited by a series of generators with $h=2,3,\dots$ and weight $T$ and another series of generators with $h=3,4,5,\dots$ and weight $T^2$. In the same spirit, we found 
\ba
\lim_{m\rightarrow\infty}I^{(A_3,A_{m-1})}=\frac{1}{(Tq^2;q)(T^2q^3;q)(T^3q^4;q)},
\ea
which means we should count the $W^{(4)}$ generators in the $\cW_4$-algebra with weight $T^3$. More generally, following from Macdonald's conjecture on the normalization of Macdonald polynomials, (\ref{Macdonald-emt-norm}), one obtain for ${\rm gcd}(n,m)=1$, 
\ba
\lim_{m\rightarrow\infty}I^{(A_{n-1},A_{m-1})}=\frac{(t;q)^{n-1}}{(q;q)^{n-1}}\frac{1}{\prod_{i=2}^{n}(t^i;q)}\frac{(q;q)^{n-1}}{(t;q)^{n-1}}\prod_{i=2}^n\frac{(t^i;q)}{(t^{i-1}q;q)}=\prod_{i=2}^n\frac{1}{(T^{i-1}q^i;q)},
\ea
which reproduces the refined character of $\cW_n$ vacuum module with weight $T^{i-1}$ for each $W^{(i)}$-generator with spin $i$. We remark again that the above Macdonald indices in the large $m$ limit is computed from the (valid) expression of the wavefunction of configuration $\lambda=\emptyset$, and the other wavefunctions with $|\lambda|\geq 1$ therefore introduce null states into the spectrum. 

One can repeat the same exercise for $G=A_1$ and $s=1$ to obtain 
\ba
\lim_{m\rightarrow\infty}I^{\mathbb{S}^1}_{(A_{1},A_{m-1})}=\frac{(1-T^2 q^2)}{(Tq;q)}.
\ea
The factor $(1-T^2q^2)$ corresponds to a universal null state, 
\ba
\lt(\frac{2}{2+m}L_{-2}-L_{-1}^2\rt)\ket{h=\frac{1-m}{2(2+m)}},
\ea
and one can thus see that the Macdonald index reproduces the refined character in the large $m$ limit for $G=A_1$ and $s=1$ (directly following Song's prescription). When we go to $s=2$, the Macdonald index computed with the strip-off factor (\ref{naive-guess-strip}) in the large $m$ limit reads 
\ba
\lim_{m\rightarrow\infty}I^{\mathbb{S}^2}_{(A_{1},A_{m-1})}=\frac{1-T^2q^2}{1-Tq^2}\frac{(1-T^2 q^3)}{(Tq;q)}.
\ea
On the other hand, there is only one common null state at level three in this series, which in the large $m$ limit takes the form,  
\ba
L_{-1}^3\ket{h=-1}.
\ea
The refined character from Song's prescription is thus given by 
\ba
\lim_{m\rightarrow\infty}\chi^{(2,2+m)}_{s=2}(q,T)=\frac{(1-T^3 q^3)}{(Tq;q)}.
\ea
It is then tempting to modify the strip-off factor to a theory-dependent one by multiplying 
\ba
\Delta{\cal F}^{\{s=2\}}_{(A_1,A_\infty)}(q,t)=\frac{(1-tq)(1-t^3)}{(1-t^2)(1-t^2q)},\label{compensating-strip-off}
\ea
so that 
\ba
\lim_{m\rightarrow\infty}\chi^{(2,2+m)}_{s=2}=\Delta{\cal F}^{\{s=2\}}_{(A_1,A_\infty)}(q,t)\lim_{m\rightarrow\infty}I^{\mathbb{S}^2}_{(A_{1},A_{m-1})}.
\ea
We can even proceed further to try to match the Macdonald indices for $s=2$ with the refined characters (\ref{ref-27-2}) and (\ref{ref-29-2}) for finite $m$ by multiplying an infinite-product factor\footnote{As the contribution from one single Schur operator is of the form $(1-t^aq^b)$ for some half integers $a$ and $b$, we try to express the compensating factor as a product of this form. }
\ba
\Delta'{\cal F}^{\{s=2\}}_{(A_1,A_2)}(q,t)&=&\frac{(1-T^3q^4)}{(1-T^4q^4)}\frac{(1-T^3q^5)(1-T^5q^5)^2}{(1-T^4q^5)^3}\frac{(1-T^5q^6)^5}{(1-T^4q^6)^2(1-T^6q^6)^3}\nn\\
&&\times \frac{(1-T^5q^7)^6(1-T^7q^7)^4}{(1-T^4q^7)(1-T^6q^7)^9}\frac{(1-T^5q^8)^5(1-T^7q^8)^{15}}{(1-T^4q^8)(1-T^6q^8)^{14}(1-T^8q^8)^5}\times\dots,\label{com-strip-12}\\
\Delta'{\cal F}^{\{s=2\}}_{(A_1,A_4)}(q,t)&=&\frac{1-T^4q^6}{1-T^5q^6}\frac{(1-T^4q^7)(1-T^6q^7)^2}{1-T^5q^7)^3}\frac{(1-T^6q^8)^4}{(1-T^5q^8)^2(1-T^7q^8)^2}\nn\\
&&\times \frac{(1-T^6q^9)^4(1-T^8q^9)}{(1-T^7q^9)^4(1-T^5q^9)}\times\dots,\label{com-strip-14}\\
\Delta'{\cal F}^{\{s=2\}}_{(A_1,A_6)}(q,t)&=&\frac{1-T^5q^8}{1-T^6q^8}\times \dots.\label{com-strip-16}
\ea
We first remark that all such corrections locates at the levels with null states. Secondly, let us focus on the term $(1-T^7q^8)^{15}$ in $\Delta'{\cal F}^{\{s=2\}}_{(A_1,A_2)}(q,t)$, which is used to decouple a bosonic d.o,f, in the UV Macdonald index with multiplicity 15. In the contribution of each Schur operator to the Macdonald index, the power of $T$ is given by $R+j_2-j_1$ and the power of $q$ is given by $R+j_1+j_2$. As for the term $(1-T^7q^8)^{15}$, $j_1$ of the corresponding Schur operators is fixed to $j_1=\frac{1}{2}$, and since Schur operators are highest weight states of SU(2)$_R$ and Lorentz symmetries, we obtain 8 possible combinations of $(R,j_2)$ from the constraints. However, as we have put the fugacity parameter of the SU(2) flavor symmetry in the UV theory to a specilal value, we also need to consider Schur operators with non-trivial flavor charge. The only possibility for the contribution $(1-T^7q^8)^{15}$ is a spin-$\frac{1}{2}$ representation of the SU(2) flavor symmetry, which further gives rise to 7 combinations of $(R,j_2)$. In total, it is consistent with the multiplicity 15 though, there will be a factor $(1-T^8q^9)^{-25} $ appearing at the next level if we continue the computation, and it is hard to be explained in this manner. We thus conclude that it is unlikely that the Macdonald index for (at least) the $(A_1,A_2)$ theory with $s=2$ surface operator inserted reproduces the refined character of $s=2$ module of the Lee-Yang singularity in the sense of Song's prescription. 

Let us go back to the $G=A_2$ case and the Macdonald index for $s_1=1,s_2=0$ in the large $m$ limit is given by 
\ba
\lim_{m\rightarrow \infty}I^{\mathbb{S}^{1,0}}_{(A_2,A_{m-1})}(q,t)=\frac{(1-T^3q^3)}{(Tq;q)(T^2q^3;q)}.\label{2m-large-m-s2}
\ea
There exist one (common) null state at level one, which asymptotes to\footnote{Note that $\frac{1}{\sqrt{m}}W_{-1}$ has the same order of normalization as $L_{-1}$ in the large $m$ limit.} 
\ba
\lt(L_{-1}+2\sqrt{\frac{3}{m}}W_{-1}\rt)\ket{h=-1,w\sim \frac{1}{3}\sqrt{\frac{m}{3}}},
\ea
in the large $m$ limit, and a (common, primary) null state at level two, whose asymptotic form is given by 
\ba
\lt(L_{-1}^2+2\sqrt{\frac{3}{m}}W_{-1}L_{-1}-2\sqrt{\frac{3}{m}}W_{-2}\rt)\ket{h=-1,w\sim \frac{1}{3}\sqrt{\frac{m}{3}}}.
\ea
As mentioned before 
\ba
L_{-1}^3\ket{h=-1,w\sim \frac{1}{3}\sqrt{\frac{m}{3}}},
\ea
also becomes null in the large $m$ limit in this series. One can thus see that (\ref{2m-large-m-s2}) gives the correct refined character in the sense of Song's prescription of the dual chiral algebra in the large $m$ limit. 

One can also expect the Macdonald index computed from our naive guess of the strip-off factor (\ref{strip-off-rank2}) to reproduce the refined character in the large $m$ limit up to a finite compensating factor similar to (\ref{compensating-strip-off}).

\subsection{$(A_2,A_m)$ index at finite $m$}

The index at large $m$ can be computed from only the wavefunction with label $\lambda=\emptyset$, which is a rather solid result, however, we have to speculate the form the wavefunction for $|\lambda|\geq 1$ based on its known behavior in the Schur and HL limit, when we want to go to finite $m$. Let us assume that the Macdonald index at finite $m$ reproduces the refined vacuum character at least for the first several levels (which is a relative natural assumption), and see what it says about the wavefunctions. 

The spectrum of the vacuum module of the $(p=3,q=7)$ minimal model can be worked out as 
\ba
{\rm level\ 0}:\ &&\ket{0},\nn\\
{\rm level\ 2}:\ &&L_{-2}\ket{0},\nn\\
{\rm level\ 3}:\ &&L_{-3}\ket{0},\quad W_{-3}\ket{0},\nn\\
{\rm level\ 4}:\ &&L_{-4}\ket{0},\quad L_{-2}L_{-2}\ket{0},\quad W_{-4}\ket{0},\nn\\
{\rm level\ 5}:\ &&L_{-5}\ket{0},\quad L_{-3}L_{-2}\ket{0},\quad W_{-5}\ket{0},\nn\\
{\rm level\ 6}:\ &&L_{-6}\ket{0},\quad L_{-4}L_{-2}\ket{0},\quad L_{-3}L_{-3}\ket{0}\quad W_{-6}\ket{0},\nn\\
&&L_{-2}L_{-2}L_{-2}\ket{0},\quad W_{-4}L_{-2}\ket{0},\nn\\
{\rm level\ 7}:\ &&L_{-7}\ket{0},\quad L_{-5}L_{-2}\ket{0},\quad L_{-4}L_{-3}\ket{0}\quad W_{-7}\ket{0},\nn\\
&&L_{-3}L_{-2}L_{-2}\ket{0},\quad W_{-5}L_{-2}\ket{0},\quad W_{-4}L_{-3}\ket{0},
\ea
where there is a non-trivial null state at level 5: 
\ba
\ket{I(5,0)}:=\lt(W_{-5}-\frac{7}{10}W_{-3}L_{-2}\rt)\ket{0},
\ea
which is a primary state with $h=5$ and $w=0$ further with two null states among its descendants at level two:  
\ba
\lt(W_{-2}-\frac{1}{4}W_{-1}L_{-1}\rt)\ket{I(5,0)},\quad \lt(L_{-2}-\frac{3}{35}L_{-1}^2-\frac{1}{10}W_{-1}^2\rt)\ket{I(5,0)}.
\ea
The refined character followed from Song's prescription for $\cW_3$-algebra is then given by 
\ba
\chi^{(3,7)}_{\{s\}=(0,0)}=1+Tq^2+(T+T^2)q^3+(T+2T^2)q^4+(T+2T^2)q^5+(T+3T^2+2T^3)q^6\nn\\
+(T+3T^2+3T^3)q^7+{\cal O}(q^8).\label{ref-37-00}
\ea
It agrees up to $q^7$-order with the Macdonald index for $(A_2,A_3)$ theory computed with the wavefunction 
\ba
f^{I_{3,4}}_{(2,1)}(q,t)=-\frac{t^5q^{2}(1-t)(1+t+qt)}{1-q^2t}\lt(\frac{(1-t^2q^2)(t^3q^2;q)}{(tq^3;q)}\rt)^{\frac{1}{2}},\label{wavefunction-A22-34}
\ea
i.e. 
\ba
I^{(A_2,A_3)}(q,t)=1+Tq^2+(T+T^2)q^3+(T+2T^2)q^4+(T+2T^2)q^5+(T+3T^2+2T^3)q^6\nn\\
+(T+3T^2+3T^3)q^7+(T+4T^2+5T^3+T^4)q^8+(T+4T^2+7T^3+2T^4)q^9+{\cal O}(q^{10}),\label{Mac-23-00}
\ea
and the Macdonald index obtained in \cite{ALS}. 

It is then not difficult to check that the (primary) null states appearing at level 6 of $(3,8)$ model and at level 8 of $(3,10)$ model respectively involve the states $W_{-3}^2\ket{0}$ and $W_{-3}^2L_{-2}\ket{0}$, i.e. states with largest weight of $T$ at the corresponding level. Song's prescription then suggests the wavefunctions of configuration $\lambda=(2,1)$ for $I_{3,5}$ and $I_{3,7}$ are given by 
\ba
f^{I_{3,5}}_{(2,1)}(q,t)=-\frac{t^6q^{2}(1-t)(1+q+qt)}{1-q^2t}\lt(\frac{(1-t^2q^2)(t^3q^2;q)}{(tq^3;q)}\rt)^{\frac{1}{2}},\\
f^{I_{3,7}}_{(2,1)}(q,t)=-\frac{t^7q^{3}(1-t)(1+t+qt)}{1-q^2t}\lt(\frac{(1-t^2q^2)(t^3q^2;q)}{(tq^3;q)}\rt)^{\frac{1}{2}},
\ea
where we note that only by replacing the factor $(1+t+qt)$ by $(1+q+qt)$ in the case of $m=5$, we can prevent the appearance of negative terms in the Macdonald index. 
Let us put an ansatz 
\ba
f^{I_{3,m}}_{(2,1)}(q,t)=-\frac{t^aq^{3+m-a}(1-t)(1+T^{\delta_{k,1}}q+qt)}{1-q^2t}\lt(\frac{(1-t^2q^2)(t^3q^2;q)}{(tq^3;q)}\rt)^{\frac{1}{2}},
\ea
for the general wavefunction of $I_{3,m}$ by further assuming a similar scaling behavior of the wavefunction as in the rank one case, (\ref{wave-rank-one}), where we set $k:=m\ {\rm mod}\ 3$. We can see that, 
\ba
C^{-1}_{(2,1)}f^{I_{3,m}}_{(2,1)}(q,t)=-T^aq^{3+m}\lt(1+(1+T^{\delta_{k,1}})q+(1+T+T^{\delta_{k,1}})q^2+{\cal O}(q^3)\rt),
\ea
which can be interpreted in the following way: the term $(1+T^{\delta_{k,1}})q$ is generated from the null state by acting $L_{-1}$ and $W_{-1}$ on it. As $L_{-1}$ vanishes when acting directly on $\ket{0}$, it only changes the label of generators in the null state, and does not change the weight of $T$. $W_{-1}$ can convert Virasoro generators to $W$-generators, so it might induce a state with weight increased by $T$ or does not change the weight, depending on the expression of the null state we start with. At level two of the null state, there are two null actions, i.e. 
\ba
\lt(W_{-2}-\frac{2}{h+3}W_{-1}L_{-1}\rt)\ket{h,0},\quad \lt(L_{-2}+\frac{3(c-2h)}{8h(5h-2)}L_{-1}^2-\frac{12(16h^2+2ch-10h+c)}{(5h-2)h(32h+2-c)}W_{-1}^2\rt)\ket{h,0},\nn\\
\ea
for $c=2-\frac{8m^2}{3+m}$ and $h=1+m$. Therefore, following Song's prescription, we see that we are left with $L_{-1}^2$ and $W_{-2}$ that generate the contribution $(1+T^{\delta_{k,1}})q^2$, and as $L_{-2}$ does not annihilate the vacuum, it raises the weight of $T$ and contributes $Tq^2$. It also explains why we have to replace the factor $(1+t+qt)$ in the wavefunction with $(1+q+qt)$ in (\ref{wavefunction-A22-32}), as there is essentially no $W$-generator in the $(3,5)$ model. We thus see that our ansatz seems to be reasonable. 

The power of $a$ can be worked out from the large $m$ spectrum (\ref{large-m-2}) and our assumption that the correction due to the wavefunction $f^{I_{3,m}}_{(2,1)}$ deletes the state with largest weight of $T$ at level $h=1+m$, 
\ba
a=2j+k+2,\quad {\rm for}\ m=3j+k,\ k=1,2.
\ea
In summary, we obtain 
\ba
f^{I_{3,3j+k}}_{(2,1)}(q,t)=-\frac{t^{2j+k+2}q^{1+j}(1-t)(1+(t/q)^{\delta_{k,1}}q+qt)}{1-q^2t}\lt(\frac{(1-t^2q^2)(t^3q^2;q)}{(tq^3;q)}\rt)^{\frac{1}{2}},
\ea
as the most natural conjecture for the general wavefunction of $I_{3,m}$. 

One can then compute the Macdonald index for $(A_2,A_3)$ theory with $s_1=1,s_2=0$ surface operator with the wavefunction (\ref{wavefunction-A22-34}) to obtain, 
\ba
I^{\mathbb{S}^{1,0}}_{(A_2,A_3)}(q,t)=1+Tq+(T+T^2)q^2+(T+2T^2)q^3+(T+3T^2+T^3)q^4+(T+3T^2+3T^3)q^5\nn\\
+(T+4T^2+5T^3+T^4)q^6+(T+4T^2+7T^3+2T^4)q^7+{\cal O}(q^8),\label{Mac-23-10}
\ea
where we used 
\ba
P_{(2,1)}(tq^{\frac{2}{3}},tq^{\frac{1}{3}})=\frac{(1+t)(1-q^3t^5+qt(1-t^2)+q^2t^2(1-t^2))}{qt^2(1-qt^2)}.
\ea
In addition to the common (primary) null states in the series of $(3,q)$ at level one and level two known as 
\ba
\lt(W_{-1}+\frac{1}{\sqrt{21}}L_{-1}\rt)\ket{(1,0)},\quad \lt(W_{-2}+\frac{2}{\sqrt{21}}L_{-2}-\frac{7}{\sqrt{21}}L_{-1}^2\rt)\ket{(1,0)},
\ea
where we denoted the highest weight state as 
\ba
\ket{(1,0)}:=\ket{h=-\frac{3}{7},w=\frac{2}{7\sqrt{21}}},
\ea
there is a null state at level four whose one descendant at level 6 is null again. We clearly see the agreement up to level 5 and since it is tricky to deal with the newly appeared (primary) null state who is generated from another null state, we will leave the comparison with the refined character at level 6 and 7 to a future work. 

\subsection{Improved POSET method}

Now we go back to the POSET approach to consider how to mend it for the rank-two case. The reason that we are in particular interested in the POSET method is that it is derived from the SH$^c$ algebra \cite{SHc}, which is essentially a rewriting of the $\mathfrak{u}(1)\times \cW_\infty$ algebra on its AFLT basis \cite{AFLT}, and studying its refined character tells us how to decouple the $\mathfrak{u}(1)$ in higher-rank cases. In the rank one case, one might also expect a correspondence between ``$>$" signs and single-trace Schur operators. The partial orderings then give us the selection rules in the Schur sector. By extending the method to higher rank cases, we will be allowed to investigate on this interpretation in the future (ref. \cite{stress-OPE, ALS, KN, Pan:2019bor}). 

In the large $m$ limit of $(A_2,A_{m-1})$ theory, we add one rule\footnote{The canonical ordering here is fixed to $(0,-3,-6,-9,-12,-15,\dots)\cup(-m,-m-3,-m-6,\dots)\cup(-2m,-2m-3,-2m-6,\dots)$. Refer to Appendix \ref{a:3m-POSET} for more details.} to the POSET method in rank one to recover the refined character up to ${\cal O}(q^{9})$. 
\begin{itemize}
\item When there are two or more $>$ appearing in an ordering and $-2m$ is before $-(m+3)$, we raise the weight of $T$ for this ordering  by one from the usual counting. 
\end{itemize}
For example, the following ordering in the usual counting has weight $T^2$, 
\ba
(0,-m_>,-3,-2m_>,-6,-9,-12,-15,\dots)\mapsto T^2q^6,
\ea
but in our improved method to carry out the refined character, it is assigned with the weight 
\ba
(0,-m_>,-3,-2m_>,-6,-9,-12,-15,\dots)\mapsto T^3q^6. 
\ea
We do not have a proof for the above improved prescription though, for example the ordering 
\ba
(0, -m_>, -3, -6, -m-3, -2m, -2m-3_>, -9,\dots ),
\ea
at level 9 is assigned with the weight $T^4q^9$, so we do not raise the weight of $T$ when $-2m-3$ is before $-m-6$ in the ordering etc. The physical meaning for our prescription is also unknown. Similarly, we checked that in the large $m$ of $(A_3,A_{m-1})$, with the rule 
\begin{itemize}
\item When there are two or more $>$ appearing in an ordering, and $-2m$ is before $-(m+4)$ or $-3m$ is before $-2m-4$, we raise the weight of $T$ for this ordering  by one from the usual counting, when we have both $-2m$ before $-(m+4)$ and $-3m$ before $-2m-4$, we still raise the weight of $T$ by one;
\end{itemize}
we managed to reproduce the refined character up to order ${\cal O}(q^8)$ from this improved POSET method for rank 3. 

In the case of finite $m$, since we have less states than the vacuum module in the large $m$ limit, we can establish a one-to-one correspondence between orderings. For example in $(3,7)$ model, the ordering 
\ba
(0,-4,-8_>,-3,-6,-7,-9,-10,-11,\dots),
\ea
at level three corresponds to 
\ba
(0,-m,-2m_>,-3,-6,-9,\dots),
\ea
in the large $m$ limit. At level four we establish 
\ba
&&(0,-3,-4,-7_>,-6,-8,-9,-10,\dots)\leftrightarrow (0,-3,-6,-m_>,-9,-12,\dots),\\
&&(0,-3,-4,-8_>,-6,-7,-9,-10,\dots)\leftrightarrow (0,-3,-m,-2m_>,-6,-9,-12,\dots),\\
&&(0,-3,-6,-9_>,-4,-7,-8,-10,\dots)\leftrightarrow (0,-3,-m,-m-3_>,-9,-12,\dots).
\ea
The spirit here is to care less about the concrete numbers but to consider the way to modify from the ``canonical" ordering at each level with weight $T$ to other orderings\footnote{In other words, we still do not have a rigorous formulation that can be stated in languages so far.}. We assign the weight to orderings in finite $m$ models same as the corresponding weight in the large $m$ limit, then this improved method of POSET leads to the refined character in the case of $(3,7)$ vacuum module, 
\ba
\chi^{POSET}_{(3,7)_{0,0}}=1+Tq^2+(T+T^2)q^3+(T+2T^2)q^4+(T+2T^2)q^5+(T+3T^2+2T^3)q^6\nn\\
+(T+3T^2+3T^3)q^7++(T+4T^2+5T^3+T^4)q^8+{\cal O}(q^9),
\ea
which agrees with the refined character (\ref{ref-37-00}) up to ${\cal O}(q^7)$ and the Macdonald index (\ref{Mac-23-00}) up to ${\cal O}(q^8)$. We mapped 
\ba
(0,-4,-8_>,-3,-7_>,-6,-9,-10,-11,-12,\dots)\leftrightarrow (0,-3,-m_>,-6,-2m_>,-9,-12,\dots),\nn\\
\ea
in the calculation, even though it will look more similar to 
\ba
(0,-m,-2m_>,-3,-m-3_>,-6,-9,\dots),
\ea
if we only look at the numbers appeared. The contribution assigned to each ordering in this case is listed in Appendix \ref{a:37-POSET}. 

The improved method also works for non-vacuum modules. For $s_1=1$, $s_2=0$ in the large $m$ limit, ${\cal X}=\{-2+3i\}_{i\in\mathbb{Z}_{\geq 0}}\cup \{-m+1+3i\}_{i\in\mathbb{Z}_{\geq 0}}\cup  \{-2m+1+3i\}_{i\in\mathbb{Z}_{\geq 0}}$, and we replace the rule for the vacuum module to ``when $-2m+1$ is before $-m-2$ in any ordering with two or more $>$, the weight of $T$ for that ordering is raised by 1". We found the agreement up to $\cO(q^6)$. In the case of $(3,7)$ model, we also established a one-to-one correspondence with the orderings in the large $m$ limit, and managed to match the result with the Macdonald index (\ref{Mac-23-10}) up to $\cO(q^6)$, where at level $6$, we mapped 
\ba
(-3_>,-2,-6,-7,-10_>,-5,-8,-9,-11,\dots)\leftrightarrow (-m+1_>,-2,-5,-m-2,-2m+1_>,-8,-11,\dots).\nn\\
\ea

\section{Fusion Rules}\label{s:fusion}

We only focus on the case of rank one, i.e. $G=A_1$, in this section. Back in the Schur limit, as 
\ba
\frac{\chi_\lambda(q^{\frac{s+1}{2}})}{\chi_\lambda(q^{\frac{1}{2}})}=\frac{q^{\frac{(s+1)\lambda}{2}}+q^{(s+1)\frac{\lambda}{2}-(s+1)}+\dots+q^{-(s+1)\frac{\lambda}{2}+(s+1)}+q^{-(s+1)\frac{\lambda}{2}}}{q^{\frac{\lambda}{2}}+q^{\frac{\lambda}{2}-1}+\dots+q^{-\frac{\lambda}{2}+1}+q^{-\frac{\lambda}{2}}}\nn\\
=q^{-s\frac{\lambda}{2}}\frac{1-q^{(\lambda+1)(s+1)}}{1-q^{s+1}}\frac{1-q}{1-q^{(\lambda+1)}},
\ea
and we will have a factor $\frac{1-q^{s+1}}{1-q}$ when converting $f_\lambda(z)$ to $C^{-1}_\lambda$, one can interpret the part 
\ba
\mathfrak{S}^s_\lambda:=q^{-s\frac{(\lambda+1)}{2}}\frac{1-q^{(\lambda+1)(s+1)}}{1-q^{(\lambda+1)}}=\chi_{s}(q^{\frac{\lambda+1}{2}}),\label{wave-surface-schur}
\ea
as the ``wavefunction" of the inserted surface operator. As shown in \cite{NSZ}, these surface operators have a clear one-to-one correspondence with primary operators (non-singular modules) in minimal models when we consider $(A_1,A_{2k})$ theories. We can then consider the situation that we add two regular punctures to the original theory, and generate two surface operators at the origin of the chiral algebra plane via the Higgsing method. These two surface operators, in the dual chiral algebra picture, are expected to fuse into a linear combination of a bunch of surface operators. The fusion rule for $\mathfrak{S}^s_\lambda$ reads 
\ba
\mathfrak{S}^{s_1}_\lambda\times\mathfrak{S}^{s_2}_\lambda=\sum_{|s_1-s_2|\leq s_3\leq s_1+s_2\ {\rm mod}\ 2}\mathfrak{S}^{s_3}_\lambda.
\ea
The above fusion rule together with the periodicity\footnote{Note that the Schur index for surface operator with $s=q-p+2+i$ needs to be interpreted to be the minus of that of $s=i$.} of the Schur index reproduce the fusion rule of Virasoro minimal models. 

It is then curious if it is possible to establish the same fusion rule in the Macdonald limit. Similar to the Schur polynomial, we have the following formula \cite{Macdonald-book}, 
\ba
\frac{P_\lambda(q^{\mu_1}t^{n-1},q^{\mu_2}t^{n-2},\dots,q^{\mu_n};q,t)}{P_\lambda(t^{n-1},t^{n-2},\dots,1;q,t)}=\frac{P_\mu(q^{\lambda_1}t^{n-1},q^{\lambda_2}t^{n-2},\dots,q^{\lambda_n};q,t)}{P_\mu(t^{n-1},t^{n-2},\dots,1;q,t)}.
\ea
In the case of $G=A_1$, we have 
\ba
\frac{P_\lambda(t^{\frac{1}{2}}q^{\frac{s}{2}},t^{-\frac{1}{2}}q^{-\frac{s}{2}};q,t)}{P_\lambda(t^{\frac{1}{2}},t^{-\frac{1}{2}};q,t)}=q^{-\frac{s}{2}}\frac{P_\lambda(tq^{s},1;q,t)}{P_\lambda(t,1;q,t)}=q^{-\frac{s}{2}}\frac{P_{(s)}(tq^{\lambda},1;q,t)}{P_{(s)}(t,1;q,t)}.
\ea
With 
\ba
P_{(s)}(t,q;q,t)=\frac{(t^2;q)_s}{(t;q)_s},
\ea
we obtain the ``wavefunction" of the surface operator (under the strip-off factor (\ref{naive-guess-strip})) as 
\ba
\mathfrak{S}^s_\lambda(q,t)=P_{(s)}(t^{\frac{1}{2}}q^{\frac{\lambda}{2}}),
\ea
which is a natural generalization of (\ref{wave-surface-schur}) to the Macdonald case. One can expand again the product of Macdonald polynomials in terms of Macdonald polynomials, 
\ba
P_\lambda(x;q,t)P_\mu(x;q,t)=\sum_{\nu}f^{\nu}{}_{\lambda\mu}P_\nu(x;q,t).
\ea
We have already seen the lost of periodicity in the Macdonald limit, and it is also not reasonable to modify the strip-off factor with (\ref{com-strip-12}) to (\ref{com-strip-16}) to restore the periodicity, however, the difference between indices for surface operators that correspond to the same module is expected to occur only at where null states exist. If this expectation is indeed true, we will see that up to the contributions from null states, we could recover the fusion rule of Virasoro minimal models in the Macdonald limit. In the large $k$ limit of $(A_1,A_{2k})$ theories, as we do not even need to worry about the periodicity, we have the correct fusion rule and it is allowed to have Macdonald indices exactly equal to the corresponding refined characters. 

\section{Conclusion}

In this article, we tested the proposal of Song's in \cite{Song} that the Macdonald index can be interpreted as a refined character of the corresponding chiral algebra with the refinement parameter $T=t/q$ counting the number of ``basic" generators in each state, for simple examples in $(A_1,A_1)$ and $(A_{n-1},A_{m-1})$ theories with ${\rm gcd}(n,m)=1$. The Macdonald index with surface operator inserted is computed via the Higgsing method used in \cite{Higgsing,NSZ}, and its dual is non-vacuum modules labeled by the same set of integer numbers as the surface operator. We compared the results obtained for the Macdonald indices and the corresponding refined characters from Song's prescription, and found both supporting and negating evidence for the proposal. As for supportive ones, we have 
\begin{itemize}
\item The refined characters of the vacuum ($s=0$) module and $s=1$ module perfectly match with the corresponding Macdonald indices for any theory in the series $(A_1,A_{2k})$ up to high orders. 

\item In $(A_1,A_1)$ theory, we have explicitly seen that the Macdonald indices with $s=0$, $s=1$ and $s=2$ surface operator respectively can be interpreted as the refined characters of the vacuum module and those generated from it by spectral flow. In particular, for $s=2$, the Macdonald index matches with the sum of three modules obtained from the Ramond sector of the $\beta\gamma$-system via spectral flow by $1/2$. 

\item The Macdonald index of $(A_{n-1},A_{m-1})$ theories with ${\rm gcd}(n,m)=1$ in the large $m$ limit reproduces the refined character of the corresponding module (up to a potential mild and finite compensation factor such as (\ref{compensating-strip-off}) to the strip-off factor), if we assign a weight $T^{j-1}$ to each spin-$j$ generator in the $\cW_n$-algebra. 
\end{itemize}

The Macdonald index in the large $m$ limit can be computed only with the wavefunction of $I_{n,m}$ with configuration $\lambda=\emptyset$ and the strip-off factor (\ref{naive-guess-strip}). Based on the assumption that the refined character tells us the Macdonald index at first several levels at finite $m$, we further discussed the expression of the wavefunction at finite $m$ for $\lambda=(2,1)$ and how to improve the POSET method. We found a match between the refined character from the POSET method and the Macdonald index up to non-trivial high orders. 

As for negating results, we have 
\begin{itemize}
\item In the computation of Macdonald index in $(A_1,A_{2k})$ theory with $s\geq2$ surface operator, we see that the results obviously cannot be interpreted as refined characters, and the periodicity of modules under the label $s$ is lost in the Macdonald index. 

\item In particular in the case of $s=2$, we pushed ahead to modify the strip-off factor (in the form of contributions from Schur operators) so that the Macdonald index matches with the refined character. It turned out an infinite product is necessary as the compensating factor to the strip-off factor. In the $(A_1,A_2)$ theory, we explicitly see that the multiplicity of factors at each level increases rapidly, and soon we found factors with too large multiplicity to be interpreted as the number of different combinations of quantum numbers giving the same Macdonald index contribution. It seems natural to expect the same phenomenon for the whole series of $(A_1,A_{2k})$ series, and even for complicated modules in higher rank theories. 

\end{itemize}

However, we found purely from our experience of studying Macdonald indices that the discrepancy with the refined character obtained from Song's prescription (for higher-rank cases with an assignment of weight $T^{j-1}$ to spin-$j$ generators) always happens at where there is a null state. This may imply that we need to modify Song's prescription to deal with null states or there are contributions from null states that are mixed in the Macdonald index with $s\geq 2$ surface operators. This will explain our success in $(A_1,A_1)$ theory and the large $m$ limit of $(A_{n-1},A_{m-1})$ theories with ${\rm gcd}(n,m)=1$, where we have either no null states or very controlled spectrum of null states. 

As this is only a primitive work in the simplest cases of chiral algebras, we would like to extend it to a more complete test to identify what went wrong in our analysis in the future. One direction to look into  is to consider the limit $t=qp^{N+1}$ and $p\rightarrow \exp(2\pi i)$ in the TQFT picture, which is also discussed to be related to the chiral algebra in \cite{MacNlimit}. 

The Macdonald index that corresponds to non-vacuum module can be obtained for theories with flavor symmetry  from an alternative way, the lens space index \cite{lens-space}. It will be be very interesting and important to repeat what we did in this article for $(A_{n-1},D_{m-1})$ AD theories, which is also important for the study of what happens in the quantum Drinfeld-Sokolov reduction in  non-vacuum modules of $\cW$-algebras from the TQFT approach. 

A huge recent breakthrough upon this topic is the discovery of a systematic way to construct the chiral algebra from 4d $\cN=2$ gauge theories via the $\Omega$-deformation \cite{Oh-Yagi,Jeong-Omega}. A similar construction was also provided in \cite{Dedushenko:2019yiw} by directly working on $S^3 \times S^1$ background, where this background can locally be approximated by the $\Omega$-background. This new approach can even allow us to consider 4d gauge theories without superconformal symmetry, by inserting a U(1)$_r$-anomalous surface operator to compensate the bulk U(1)$_r$ anomaly in the spirit of anomaly inflow \cite{Callan-Harvey}. It may enable us to consider the chiral algebra for simple but Lagrangian theories such as the pure $\cN=2$ Yang-Mills in the future. 

An $\cN=1$ Lagrangian approach to the Argyres-Douglas theories has been developed since \cite{Maruyoshi-Song, Maruyoshi-Song-JHEP, Agarwal-Maruyoshi-Song}. In this approach, one starts from a normal 4d $\cN=1$ gauge theory with Lagrangian description, and surprisingly some special RG flow takes us to IR SCFT with enhanced supersymmetry, such as the AD theories. Inserting surface operators in this context will be a very interesting and robust direction to work on as a future work. 


\section*{Acknowledgement}

We would like to thank O. Foda, M. Fukuda, S. Nakamura, S. Nawata, T Nishinaka, J. Song, M. Dedushenko for inspiring and helpful discussions and comments on the article. The work of AW was partially supported by the Program for Leading
Graduate Schools, MEXT, Japan.

\appendix

\section{Macdonald Polynoials}\label{a:Macdonald}

We summarize the properties of $A_1$ and $A_2$ Macdonald polynomials in this Appendix mainly based on \cite{Macdonald-book}. 

Let us first recall the formal expansion series of a general $A$-type Macdonald polynomial, which can be found in \cite{Mac-Shiraishi}. With the notation 
\ba
c_n(\theta;s|q,t)&=&\prod_{k=2}^n\prod_{1\leq i<j\leq k}\frac{(q^{\sum_{a=k+1}^n\theta_{ia}-\theta_{ja}}ts_j/s_i;q)_{\theta_{ik}}}{(q^{\sum_{a=k+1}^n\theta_{ia}-\theta_{ja}}qs_j/s_i;q)_{\theta_{ik}}}\nn\\
&&\times \prod_{k=2}^n\prod_{1\leq i\leq j< k}\frac{(q^{-\theta_{jk}+\sum_{a=k+1}^n\theta_{ia}-\theta_{ja}}qs_j/ts_i;q)_{\theta_{ik}}}{(q^{-\theta_{jk}+\sum_{a=k+1}^n\theta_{ia}-\theta_{ja}}s_j/s_i;q)_{\theta_{ik}}},
\ea
where $\theta=(\theta_{ij})_{i,j=1}^n$ is an upper triangular matrix with zero diagonal entires, i.e. satisfying $\theta_{ij}=0$ for $j\leq i$, the pre-Macdonald fractional function is given by 
\ba
p_n(x;s|q,t)=\sum_\theta c_n(\theta;s|q,t)\prod_{1\leq i <j\leq n}\lt(x_j/x_i\rt)^{\theta_{ij}}.
\ea
We will mainly focus on the case $n=2$ and $n=3$ in this article, so let us write down a more concrete expression of the coefficient $c_n(\theta;s|q,t)$ in these cases. For $n=2$, $\theta$ has only one non-vanishing entry, $\theta_{12}$, 
\ba
c_2(\theta_{12};s_1,s_2|q,t)=\frac{(ts_2/s_1;q)_{\theta_{12}}}{(qs_2/s_1;q)_{\theta_{12}}}\frac{(t;q)_{\theta_{12}}}{(q;q)_{\theta_{12}}}\lt(\frac{q}{t}\rt)^{\theta_{12}}.
\ea
For $n=3$, we have 
\ba
c_3(\theta;s_1,s_2,s_3|q,t)=\frac{(q^{\theta_{13}-\theta_{23}}ts_2/s_1;q)_{\theta_{12}}}{(q^{\theta_{13}-\theta_{23}+1}s_2/s_1;q)_{\theta_{12}}}\frac{(t;q)_{\theta_{12}}}{(q;q)_{\theta_{12}}}\frac{(ts_2/s_1;q)_{\theta_{13}}}{(qs_2/s_1;q)_{\theta_{13}}}\frac{(ts_3/s_1;q)_{\theta_{13}}}{(qs_3/s_1;q)_{\theta_{13}}}\frac{(ts_3/s_2;q)_{\theta_{23}}}{(qs_3/s_2;q)_{\theta_{23}}}\nn\\
\times \frac{(t;q)_{\theta_{13}}}{(q;q)_{\theta_{13}}}\frac{(t;q)_{\theta_{23}}}{(q;q)_{\theta_{23}}}\frac{(q^{-\theta_{23}+1}s_2/ts_1;q)_{\theta_{13}}}{(q^{-\theta_{23}}s_2/s_1;q)_{\theta_{13}}}\lt(\frac{q}{t}\rt)^{\theta_{12}+\theta_{13}+\theta_{23}}.
\ea
The (unnormalized) Macdonald polynomial is given by 
\ba
P_\lambda(x;s|q,t)=\lt(\prod_{i=1}^n x_i^{\lambda_i}\rt)p_n(x;s|q,t),
\ea
where the label $\lambda=\{\lambda_1,\lambda_2,\dots,\lambda_n\}$ is a Young diagram, i.e. satisfying $\lambda_1\geq \lambda_2\geq \lambda_3\geq \dots\geq \lambda_n\geq 0$. 

Parameters $s=\{s_i\}_{i=1}^n$ can be fixed by substituting the formal expansion series into the eigenvalue equation, 
\ba
DP_\lambda(x;s|q,t)=\epsilon(s)P_\lambda(x;s|q,t),
\ea
where $\epsilon$ is some eigenvalue that can be uniquely determined by $s$, and $D$ is the Macdonald operator, 
\ba
D=\sum_{i=1}^n\prod_{j\neq i}\frac{tx_i-x_j}{x_i-x_j}T_{q,i},
\ea
with $T_{q,i}$ the shift operator for the $i$-th variable $x_i$ by $q$, i.e. $T_{q,i}x_j=q^{\delta_{i,j}}x_j$. For $n=2$, by putting $z:=x_2/x_1$, we have 
\ba
\sum_{\theta_{12}}\frac{t-z}{1-z}c_2(\theta_{12};s|q,t)q^{\lambda_1}(q^{-1}z)^{\theta_{12}}+\frac{1-tz}{1-z}c_2(\theta_{12};s|q,t)q^{\lambda_2}(qz)^{\theta_{12}}=\sum_{\theta_{12}}\epsilon(s)c_2(\theta_{12};s|q,t)z^{\theta_{12}},
\ea
which leads to 
\ba
\lt(\frac{tq^{\lambda_1}}{q^{\theta_{12}+1}}+q^{\lambda_2+\theta_{12}+1}-\epsilon(s)\rt)c_2(\theta_{12}+1;s|q,t)=\lt(q^{\lambda_1-\theta_{12}}+tq^{\lambda_2+\theta_{12}}-\epsilon(s)\rt)c_2(\theta_{12};s|q,t),
\ea
and agrees with the recursive relation of $c_2(\theta_{12};s|q,t)$ for 
\ba
\epsilon(s)=s_1+s_2,\quad s_1=tq^{\lambda_1},\quad s_2=q^{\lambda_2}.
\ea
Note that $(ts_2/s_1;q)_{\theta_{12}}$ vanishes when $\theta_{12}\geq \lambda_1-\lambda_2+1$, and it is clear that $P_\lambda(x_1,x_2;s|q,t)$ is a polynomial symmetric about $x_1\leftrightarrow x_2$. 

In the same way, one can easily find that for $n=3$, 
\ba
\epsilon(s)=s_1+s_2+s_3,\quad s_1=t^2q^{\lambda_1},\quad s_2=tq^{\lambda_2},\quad s_3=q^{\lambda_3}.
\ea

Note that we can get rid of the ``U(1)" part of the Macdonald polynomial by putting $\lambda_{n}=0$. For example, for $n=2$, the Macdonald polynomial can be rewritten as (with the redefinition $\lambda=\lambda_1$) 
\ba
P_\lambda(x|q,t)=x_1^{\lambda}\sum_{\theta=0}^{\lambda}\frac{(q^{\lambda-\theta+1};q)_\theta (t;q)_\theta}{(tq^{\lambda-\theta};q)_\theta (q;q)_\theta}z^{\theta}.\label{n=2Mac}
\ea
By further putting $x_1^{-1}=x_2=\zeta$ and multiplying $\frac{(t;q)_\lambda}{(q;q)_\lambda}$, we obtain a more familiar expression of the Macdonald polynomial, 
\ba
P'_\lambda(\zeta|q,t)=\sum_{\theta=0}^\lambda \frac{(t;q)_{\lambda-\theta}(t;q)_\theta}{(q;q)_{\lambda-\theta}(q;q)_\theta}\zeta^{2\theta-\lambda}.
\ea
For $n=3$, we only write down the explicit form of the $A_2$ Macdonald polynomials for the first several Young diagrams: 
\ba
&&P_\emptyset(x|q,t)=1,\quad P_{(1,0)}(x|q,t)=x_1+x_2+x_3,\\
&&P_{(1,1)}(x|q,t)=x_1x_2+x_1x_3+x_2x_3,\\
&&P_{(2,0)}(x|q,t)=x_1^2+x_2^2+x_3^2+\frac{(1+q)(1-t)}{1-qt}(x_1x_2+x_1x_3+x_2x_3),\\
&&P_{(2,1)}(x|q,t)=x_1^2x_2+x_2^2x_1+x_3^2x_1+x_1^2x_3+x_3^2x_2+x_2^2x_3+\lt(\frac{(1+q)(1-t)}{1-qt}+\frac{(1-t^2)(1-tq^2)}{(1-qt)(1-qt^2)}\rt)x_1x_2x_3,\nn\\
&&\\
&&P_{(3,0)}(x|q,t)=x_1^3+x_2^3+x_3^3+\frac{(1+q+q^2)(1-t)}{1-q^2t}\lt(x_1^2x_2+x_2^2x_1+x_1^2x_3+x_2^2x_3+x_3^2x_1+x_3^2x_2\rt)\nn\\
&&+\frac{(1+q)(1+q+q^2)(1-t)^2}{(1-qt)(1-q^2t)}x_1x_2x_3,\\
&&P_{(3,3)}(x|q,t)=x_1^3x_2^3+x_1^3x_3^3+x_2^3x_3^3+\frac{(1-t)(1+q+q^2)}{1-q^2t}\lt(x^3_1x_2^2x_3+x_2^3x_1^2x_3+x_1^3x_3^2x_2+x_2^3x_3^2x_1\rt.\nn\\
&&\lt.+x_3^3x_1^2x_2+x_3^3x_2^2x_1\rt)+\frac{(1+q)(1+q+q^2)(1-t)^2}{(1-qt)(1-q^2t)}x_1^2x_2^2x_3^2,
\ea
We need to set $x_3=x_1^{-1}x_2^{-1}$ to decouple the U(1) part of the $A_2$ Macdonald polynomials in this article. In particular, by doing so, we can re-expand the Macdonald polynomials in terms of the characters of SU(3), $\chi^{su(3)}_\lambda(z)$, with $x_1=z_1$, $x_2=z_2/z_1$, and for example, we have 
\ba
&&P_{(2,1)}(x|q,t)=\chi^{su(3)}_{(2,1)}(z)+\frac{(q-t)(1+t)}{1-qt^2}\chi^{su(3)}_\emptyset(z),\\
&&P_{(3,0)}(x|q,t)=\chi^{su(3)}_{(3,0)}(z)+\frac{(q-t)(1+q)}{1-q^2t}\chi^{su(3)}_{(2,1)}(z)+\frac{(q-t)(q^2-t)}{(1-qt)(1-q^2t)}\chi^{su(3)}_\emptyset(z),\\
&&P_{(3,3)}(x|q,t)=\chi^{su(3)}_{(3,3)}(z)+\frac{(q-t)(1+q)}{1-q^2t}\chi^{su(3)}_{(2,1)}(z)+\frac{(q-t)(q^2-t)}{(1-qt)(1-q^2t)}\chi^{su(3)}_\emptyset(z).
\ea

The Pieri rule is also a very important property for the Macdonald polynomials. 
\ba
P_\lambda(x|q,t)e_r(x)=\sum_\mu B_{\mu/\lambda}(q,t)P_\mu(x|q,t),
\ea
where $\mu$ runs over all Young diagrams that can be obtained from $\lambda$ plus an $r$-vertical strip, 
\ba
e_r(x)=\sum_{i_1<i_2<\dots<i_r}x_{i_1}x_{i_2}\dots x_{i_r},
\ea
and 
\ba
B_{\mu/\lambda}(q,t)=\prod_{\substack{1\leq i<j\leq n\\\lambda_i=\mu_i,\lambda_j+1=\mu_j}}\frac{1-q^{\lambda_i-\lambda_j}t^{j-i-1}}{1-q^{\lambda_i-\lambda_j}t^{j-i}}\frac{1-q^{\mu_i-\mu_j}t^{j-i+1}}{1-q^{\mu_i-\mu_j}t^{j-i}}.
\ea
Note that $e_r(x^{-1})e_n(x)=e_{n-r}(x)$. 
For $n=2$, $r=1$, we have 
\ba
P_\lambda(x|q,t)e_1(x)=P_{\lambda+1}(x|q,t)+\frac{(1-q^{\lambda})(1-q^{\lambda-1} t^2)}{(1-q^{\lambda}t)(1-q^{\lambda-1} t)}x_1x_2P_{\lambda-1}(x|q,t),
\ea
where $\lambda=\lambda_1$ is a one-raw Young diagram and we used $x_1x_2P_{(\lambda,0)}(x|q,t)=P_{(\lambda+1,1)}(x|q,t)$. For $n=3$, $r=1$, 
\ba
P_{(\lambda_1,\lambda_2)}(x|q,t)e_1(x)=P_{(\lambda_1+1,\lambda_2)}(x|q,t)+\frac{(1-q^{\lambda_1-\lambda_2})(1-q^{\lambda_1-\lambda_2-1} t^2)}{(1-q^{\lambda_1-\lambda_2}t)(1-q^{\lambda_1-\lambda_2-1} t)}P_{(\lambda_1,\lambda_2+1)}(x|q,t)\nn\\
+\frac{(1-q^{\lambda_1}t)(1-q^{\lambda_1-1} t^3)}{(1-q^{\lambda_1}t^2)(1-q^{\lambda_1-1} t^2)}\frac{(1-q^{\lambda_2})(1-q^{\lambda_2-1} t^2)}{(1-q^{\lambda_2}t)(1-q^{\lambda_2-1} t)}x_1x_2x_3P_{(\lambda_1-1,\lambda_2-1)}(x|q,t),
\ea
and for $n=3$, $r=2$, 
\ba
P_{(\lambda_1,\lambda_2)}(x|q,t)e_2(x)=P_{(\lambda_1+1,\lambda_2+1)}(x|q,t)+\frac{(1-q^{\lambda_1-\lambda_2})(1-q^{\lambda_1-\lambda_2-1} t^2)}{(1-q^{\lambda_1-\lambda_2}t)(1-q^{\lambda_1-\lambda_2-1} t)}\frac{(1-q^{\lambda_1}t)(1-q^{\lambda_1-1} t^3)}{(1-q^{\lambda_1}t^2)(1-q^{\lambda_1-1} t^2)}\nn\\
\times x_1x_2x_3P_{(\lambda_1-1,\lambda_2)}(x|q,t)+\frac{(1-q^{\lambda_2})(1-q^{\lambda_2-1} t^2)}{(1-q^{\lambda_2}t)(1-q^{\lambda_2-1} t)}x_1x_2x_3P_{(\lambda_1,\lambda_2-1)}(x|q,t).\nn\\
\ea

Anther aspect of the Macdonald polynomials that will be extremely important in this article is their integration. In particular, we need to normalize the Macdonald polynomial with respect to the integral under the measure 
\ba
\Delta_{q,t}(\{z_i\}_{i=1}^n)=\frac{1}{n!}\prod_{i\neq j}\frac{(z_i/z_j;q)}{(t z_i/z_j;q)}.
\ea
For $n=2$, it reduces to 
\ba
\Delta_{q,t}(\zeta)=\frac{1}{2}\frac{(\zeta^2;q)(\zeta^{-2};q)}{(t\zeta^2;q)(t\zeta^{-2};q)}.
\ea
We have 
\ba
(z;q)=\sum_{n=0}^\infty (-1)^n\frac{q^{\frac{n(n-1)}{2}}}{(q;q)_n}z^n,
\ea
and 
\ba
\frac{1}{(z;q)}=\sum_{n=0}^\infty \frac{1}{(q;q)_n}z^n,
\ea
therefore we can expand 
\ba
\Delta_{q,t}(\zeta)=\frac{1}{2}\sum_{n,m,k,l=0}^\infty(-1)^{n+m}\frac{q^{\frac{n(n-1)}{2}+\frac{m(m-1)}{2}}}{(q;q)_{n}(q;q)_{m}(q;q)_{k}(q;q)_{l}}t^{k+l}\zeta^{2n-2m+2k-2l}.
\ea
The constant terms in the above expansion can be found as 
\ba
\frac{1}{2}\sum_{N=0}^\infty\sum_{n,m=0}^N (-1)^{n+m}\frac{q^{\frac{n(n-1)}{2}+\frac{m(m-1)}{2}}}{(q;q)_{n}(q;q)_{m}(q;q)_{N-n}(q;q)_{N-m}}t^{2N-n-m}\nn\\
=\frac{1}{(q;q)}-\frac{1+q}{1-q}\frac{1}{(q;q)}t+\dots
\ea
Macdonald conjectured \cite{Macdonald-book} that the above express is equal to 
\ba
\frac{(t;q)(tq;q)}{(q;q)(t^2;q)}.
\ea
His conjecture for a general $n$ is given by 
\ba
\prod_{i=1}^n\frac{(t;q)(t^{i-1}q;q)}{(q;q)(t^i;q)}.\label{Macdonald-emt-norm}
\ea
With this explicit expression for the integral of $1$, we can work out the integral of a general pair of Macdonald polynomials by using the Pieri rule. Let us introduce a short notation for the integral as an inner product, 
\ba
(P_\lambda,P_\mu)=\oint\prod_{i=1}^{n-1}\frac{{\rm d}x_i}{2\pi ix_i}\Delta_{q,t}(\{x_i\})P_\lambda(x^{-1}|q,t)P_\mu(x|q,t).
\ea
As $P_\lambda$ is the eigenstate of the Macdonald operator $D$, it is not difficult to show that 
\ba
(P_\lambda,P_\mu)\propto \delta_{\lambda\mu}.
\ea
We can then evaluate for $n=2$ 
\ba
(P_\lambda,e_1P_\mu)=(\bar{e}_1e_2P_\lambda,e_2P_\mu)=(e_1P_\lambda,e_2P_\mu),
\ea
where we use the notation $\bar{e}_1(x)=e_1(x^{-1})$. Let us set $\lambda=\mu+1$, we obtain 
\ba
(P_{\mu+1},P_{\mu+1})=\frac{(1-q^{\mu+1})(1-q^\mu t^2)}{(1-q^{\mu+1}t)(1-q^\mu t)}(P_{\mu},P_{\mu}),
\ea
and together with Macdonald's conjecture for the constant integral, 
\ba
(P_{\lambda},P_{\lambda})=\frac{(tq^\lambda;q)(tq^{\lambda+1};q)}{(q^{\lambda+1};q)(t^2q^\lambda;q)}.
\ea
For $n=3$, 
\ba
(P_\lambda,e_1P_\mu)=(\bar{e}_1e_3P_\lambda,e_3P_\mu)=(e_2P_\lambda,e_3P_\mu).
\ea
When $(\lambda_1,\lambda_2)=(\mu_1+1,\mu_2)$, we obtain from the above formula that 
\ba
(P_{(\mu_1+1,\mu_2)},P_{(\mu_1+1,\mu_2)})=\frac{(1-q^{\mu_1-\mu_2+1})(1-q^{\mu_1-\mu_2} t^2)}{(1-q^{\mu_1-\mu_2+1}t)(1-q^{\mu_1-\mu_2} t)}\frac{(1-q^{\mu_1+1}t)(1-q^{\mu_1} t^3)}{(1-q^{\mu_1+1}t^2)(1-q^{\mu_1} t^2)}(P_{(\mu_1,\mu_2)},P_{(\mu_1,\mu_2)}),\nn\\
\ea
and for $(\lambda_1,\lambda_2)=(\mu_1,\mu_2+1)$, 
\ba
\frac{(1-q^{\mu_1-\mu_2})(1-q^{\mu_1-\mu_2-1} t^2)}{(1-q^{\mu_1-\mu_2}t)(1-q^{\mu_1-\mu_2-1} t)}(P_{(\mu_1,\mu_2+1)},P_{(\mu_1,\mu_2+1)})=\frac{(1-q^{\mu_2+1})(1-q^{\mu_2} t^2)}{(1-q^{\mu_2+1}t)(1-q^{\mu_2} t)}(P_{(\mu_1,\mu_2)},P_{(\mu_1,\mu_2)}),\nn\\
\ea
or 
\ba
(P_{(\mu_1,\mu_2+1)},P_{(\mu_1,\mu_2+1)})=\frac{(1-q^{\mu_1-\mu_2}t)(1-q^{\mu_1-\mu_2-1} t)}{(1-q^{\mu_1-\mu_2})(1-q^{\mu_1-\mu_2-1} t^2)}\frac{(1-q^{\mu_2+1})(1-q^{\mu_2} t^2)}{(1-q^{\mu_2+1}t)(1-q^{\mu_2} t)}(P_{(\mu_1,\mu_2)},P_{(\mu_1,\mu_2)}).\nn\\
\ea
More explicitly, we have 
\ba
&&(P_{\emptyset},P_{\emptyset})=\frac{1}{1-t^2}\frac{(t;q)^2(tq;q)}{(q;q)^2(t^3;q)},\quad (P_{(1,0)},P_{(1,0)})=\frac{1}{(t^2;q)_2}\frac{(t;q)}{(q;q)}\frac{(tq;q)(tq;q)}{(q^2;q)(t^3q;q)},\\
&&(P_{(1,1)},P_{(1,1)})=\frac{1}{(t^2;q)_2}\frac{(t;q)}{(q;q)}\frac{(tq;q)(tq;q)}{(q^2;q)(t^3q;q)},\quad (P_{(2,0)},P_{(2,0)})=\frac{1}{(t^2;q)_3}\frac{(t;q)}{(q;q)}\frac{(tq;q)(tq^2;q)}{(q^3;q)(t^3q^2;q)},\nn\\
&&\\
&&(P_{(2,1)},P_{(2,1)})=\frac{(1-tq^2)}{(t^2q;q)_2(1-t^2q)}\frac{(tq;q)}{(q^2;q)}\frac{(tq;q)(tq^2;q)}{(q^2;q)(t^3q^2;q)},\\
&&(P_{(3,0)},P_{(3,0)})=\frac{1}{(t^2;q)_4}\frac{(t;q)}{(q;q)}\frac{(tq;q)(tq^3;q)}{(q^4;q)(t^3q^3;q)},\quad (P_{(2,2)},P_{(2,2)})=\frac{1}{(t^2;q)_3}\frac{(t;q)}{(q;q)}\frac{(tq;q)(tq^2;q)}{(q^3;q)(t^3q^2;q)},\nn\\
\\
&&(P_{(3,2)},P_{(3,2)})=\frac{1-tq^3}{(t^2q;q)_2(t^2q^2;q)_2}\frac{(tq;q)}{(q^2;q)}\frac{(tq^2;q)(tq^2;q)}{(q^3;q)(t^3q^3;q)},\\
&&(P_{(3,3)},P_{(3,3)})=\frac{1}{(t^2;q)_4}\frac{(t;q)}{(q;q)}\frac{(tq;q)(tq^3;q)}{(q^4;q)(t^3q^3;q)}.
\ea

The Pieri rule is also convenient to determine the coefficient of the Macdonald polynomial in terms of the characters. For $n=2$, we can start from $P_1(z|q,t)=\chi_1(z)$, and obtain 
\ba
&&P_2(z|q,t)=\chi_2(z)+\frac{q-t}{1-qt}\chi_0(z),\\
&&P_3(z|q,t)=\chi_3(z)+\frac{(1+q)(q-t)}{1-q^2t}\chi_1(z),\\
&&P_4(z|q,t)=\chi_4(z)+\frac{(q-t)(1+q+q^2)}{1-q^3t}\chi_2+\frac{q(1+q^2)(q-t)(1-t)}{(1-q^2t)(1-q^3t)}\chi_0(z),\\
&&P_5(z|q,t)=\chi_5(z)+\frac{(q-t)(1+q+q^2+q^3)}{1-q^4t}\chi_3(z)+\frac{q(q-t)(1-t)(1+q+q^2+q^3+q^4)}{(1-q^3t)(1-q^4t)}\chi_1(z),\nn\\
\\
&&P_6(z|q,t)=\chi_6(z)+\frac{(q-t)(1+q+q^2+q^3+q^4)}{1-q^5t}\chi_4(z)+\frac{q(q-t)(1-t)(1-q+q^2)(1+q+q^2)^2}{(1-q^4 t)(1-q^5t)}\chi_2(z)\nn\\
&&+\frac{q^2(q-t)(1-t)(1-qt)(1+q^2+q^3+q^4+q^6)}{(1-q^3t)(1-q^4t)(1-q^5t)}\chi_0(z).
\ea

At last, we comment that the following specialization holds 
\ba
P_\lambda(1,t,t^2,\dots,t^{n-1}|q,t)=t^{n(\lambda)}\prod_{(i,j)\in\lambda}\frac{1-q^{j-1}t^{n-i+1}}{1-q^{a(i,j)}t^{\ell(i,j)+1}}.\label{P-rho}
\ea

\section{Hall-Littlewood Polynomials}\label{a:HL}

In this section, we present the properties of the Hall-Littlewood polynomials as the $q\rightarrow 0$ limit of the Macdonald limit. 

For $n=2$, it is very easy to work out the $q\rightarrow 0$ limit of (\ref{n=2Mac}), 
\ba
P_{\lambda}^{HL}(\zeta|t)=\lt[\frac{\zeta^{\lambda+1}-\zeta^{-\lambda-1}}{\zeta-\zeta^{-1}}-t\frac{\zeta^{\lambda-1}-\zeta^{-\lambda+1}}{\zeta-\zeta^{-1}}\rt].
\ea
The measure is simplified to 
\ba
\Delta_t(\zeta)=\frac{1}{2}\frac{(1-\zeta^2)(1-\zeta^{-2})}{(1-t\zeta^2)(1-t\zeta^{-2})},
\ea
and the integration 
\ba
\oint\frac{{\rm d}\zeta}{2\pi i\zeta}\Delta_t(\zeta)P_{\lambda}^{HL}(\zeta|t)P_{\mu}^{HL}(\zeta^{-1}|t),
\ea
can be explicitly computed to for $\lambda,\mu\neq\emptyset$ 
\ba
\frac{1}{2}\oint\frac{\zeta{\rm d}\zeta}{2\pi i}\frac{1}{(1-t\zeta^2)(\zeta^2-t)}\lt(\zeta^{\lambda+1}-t\zeta^{\lambda-1}+t\zeta^{-\lambda+1}-\zeta^{-\lambda-1}\rt)\lt(\zeta^{\mu+1}-t\zeta^{\mu-1}+t\zeta^{-\mu+1}-\zeta^{-\mu-1}\rt)\nn\\
=-\delta_{\mu\nu},
\ea
and for $\lambda=\mu=\emptyset$
\ba
\frac{1}{2}\oint\frac{{\rm d}\zeta}{2\pi i\zeta}\frac{(1-\zeta^2)^2}{(1-t\zeta^2)(t-\zeta^{2})}=\frac{1}{1+t},\label{integral-HL-0}
\ea
where the integral contour is the unit circle surrounding poles at $\zeta=0$ and $\zeta=\pm t^{\frac{1}{2}}$. We see that the normalized Hall-Littlewood polynomial for $n=2$ is simply given by 
\ba
\bar{P}_{\lambda}^{ HL}(\zeta|t)=\lt[\frac{\zeta^{\lambda+1}-\zeta^{-\lambda-1}}{\zeta-\zeta^{-1}}-t\frac{\zeta^{\lambda-1}-\zeta^{-\lambda+1}}{\zeta-\zeta^{-1}}\rt]=\chi^{su(2)}_\lambda(\zeta)-t\chi^{su(2)}_{\lambda-2}(\zeta),
\ea
for $\lambda\neq\emptyset$, and 
\ba
\bar{P}^{HL}_\emptyset(\zeta|t)=\sqrt{1+t}.
\ea

For $n=3$, the measure is given by 
\ba
\Delta_t(z_1,z_2)=\frac{1}{6}\frac{(1-z_1/z_2)(1-z_2/z_1)(1-z_1^2z_2)(1-z_1^{-2}z^{-1}_2)(1-z_1z_2^2)(1-z_1^{-1}z_2^{-2})}{(1-tz_1/z_2)(1-tz_2/z_1)(1-tz_1^2z_2)(1-tz_1^{-2}z^{-1}_2)(1-tz_1z_2^2)(1-tz_1^{-1}z_2^{-2})}.
\ea
The integration over $1$ can be computed to 
\ba
\oint\frac{{\rm d}z_1}{2\pi iz_1}\frac{{\rm d}z_2}{2\pi iz_2}\Delta_t(z_1,z_2)=\frac{1}{(1+t)(1+t+t^2)},
\ea
which agrees with the conjecture of Macdonald's in the HL limit. One can again write down a closed-form explicit formula for Hall-Littlewood polynomials for $n=3$. When $\lambda_1>\lambda_2$, 
\ba
P^{HL}_{(\lambda_1,\lambda_2)}(x_1,x_2,x_3|t)=\sum_{\sigma\in S_3}x_{\sigma(1)}^{\lambda_1}x_{\sigma(2)}^{\lambda_2}\frac{(x_{\sigma(1)}-tx_{\sigma(2)})(x_{\sigma(1)}-tx_{\sigma(3)})(x_{\sigma(2)}-tx_{\sigma(3)})}{(x_{\sigma(1)}-x_{\sigma(2)})(x_{\sigma(1)}-x_{\sigma(3)})(x_{\sigma(2)}-x_{\sigma(3)})},
\ea
and when $\lambda_1=\lambda_2=\lambda$, we have 
\ba
P^{HL}_{(\lambda,\lambda)}(x_1,x_2,x_3|t)=\frac{1}{1+t}\sum_{\sigma\in S_3}(x_{\sigma(1)}x_{\sigma(2)})^{\lambda}\frac{(x_{\sigma(1)}-tx_{\sigma(2)})(x_{\sigma(1)}-tx_{\sigma(3)})(x_{\sigma(2)}-tx_{\sigma(3)})}{(x_{\sigma(1)}-x_{\sigma(2)})(x_{\sigma(1)}-x_{\sigma(3)})(x_{\sigma(2)}-x_{\sigma(3)})}.
\ea
The inner product in the case of Hall-Littlewood is also simplified a lot. We have for $\lambda_1>\lambda_2+1$, 
\ba
(P^{HL}_{(\lambda_1+1,\lambda_2)},P^{HL}_{(\lambda_1+1,\lambda_2)})=(P^{HL}_{(\lambda_1,\lambda_2)},P^{HL}_{(\lambda_1,\lambda_2)}),
\ea
for $\lambda_1=\lambda_2=\lambda$, 
\ba
(P^{HL}_{(\lambda+1,\lambda)},P^{HL}_{(\lambda+1,\lambda)})=(1+t)(P^{HL}_{(\lambda,\lambda)},P^{HL}_{(\lambda,\lambda)}),
\ea
for $\lambda_1>\lambda_2+1>1$, 
\ba
(P^{HL}_{(\lambda_1,\lambda_2+1)},P^{HL}_{(\lambda_1,\lambda_2+1)})=(P^{HL}_{(\lambda_1,\lambda_2)},P^{HL}_{(\lambda_1,\lambda_2)}),
\ea
and for $\lambda=\lambda_1=\lambda_2+1>1$, 
\ba
(P^{HL}_{(\lambda,\lambda)},P^{HL}_{(\lambda,\lambda)})=\frac{1}{(1+t)}(P^{HL}_{(\lambda,\lambda-1)},P^{HL}_{(\lambda,\lambda-1)}).
\ea

As 
\ba
&&(P^{HL}_{\emptyset},P^{HL}_{\emptyset})=\frac{1}{(1+t)(1+t+t^2)},\quad (P^{HL}_{(1,0)},P^{HL}_{(1,0)})=(P^{HL}_{(1,1)},P^{HL}_{(1,1)})=\frac{1}{1+t},\\
&&(P^{HL}_{(2,1)},P^{HL}_{(2,1)})=1,
\ea
we obtain for $\lambda_1>\lambda_2\geq 1$, 
\ba
(P^{HL}_{(\lambda_1,\lambda_2)},P^{HL}_{(\lambda_1,\lambda_2)})=1,
\ea
for $\lambda_2=0$, $\lambda_1\geq 1$, 
\ba
(P^{HL}_{(\lambda_1,0)},P^{HL}_{(\lambda_1,0)})=\frac{1}{1+t},
\ea
and $\lambda_1=\lambda_2=\lambda\geq 1$, 
\ba
(P^{HL}_{(\lambda,\lambda)},P^{HL}_{(\lambda,\lambda)})=\frac{1}{1+t}.
\ea

The Pieri rule also helps to write down the explicit form of the $A_2$ HL polynomials in terms of SU(3) characters, as we have 
\ba
P_{(1,0)}(x|t)=\chi^{su(3)}_{(1,0)}(z_1,z_2),\quad P_{(1,1)}(x|t)=\chi^{su(3)}_{(1,1)}(z_1,z_2).\label{HL-schur-s}
\ea
What we obtain is 
\ba
&&P_{(2,0)}(x|t)=\chi^{su(3)}_{(2,0)}(z_1,z_2)-t\chi^{su(3)}_{(1,1)}(z_1,z_2),\\
&&P_{(2,1)}(x|t)=\chi^{su(3)}_{(2,1)}(z_1,z_2)-(t+t^2)\chi^{su(3)}_{\emptyset}(z_1,z_2),\\
&&P_{(2,2)}(x|t)=\chi^{su(3)}_{(2,2)}(z_1,z_2)-t\chi^{su(3)}_{(1,0)}(z_1,z_2),\\
&&P_{(3,0)}(x|t)=\chi^{su(3)}_{(3,0)}(z_1,z_2)-t\chi^{su(3)}_{(2,1)}(z_1,z_2)+t^2\chi^{su(3)}_{\emptyset}(z_1,z_2),\\
&&P_{(3,1)}(x|t)=\chi^{su(3)}_{(3,1)}(z_1,z_2)-t\chi^{su(3)}_{(2,2)}(z_1,z_2)-t\chi^{su(3)}_{(1,0)}(z_1,z_2),\\
&&P_{(4,0)}(x|t)=\chi^{su(3)}_{(4,0)}(z_1,z_2)-t\chi^{su(3)}_{(3,1)}(z_1,z_2)+t^2\chi^{su(3)}_{(1,0)}(z_1,z_2),\\
&&P_{(3,3)}(x|t)=\chi^{su(3)}_{(3,3)}(z_1,z_2)-t\chi^{su(3)}_{(2,1)}(z_1,z_2)+t^2\chi^{su(3)}_{\emptyset}(z_1,z_2).\label{HL-schur-e}
\ea

At last, we also compute $P_\lambda(1,t,t^2|q,t)$ for $\lambda=(\lambda_1,\lambda_2)$. When $\lambda_2=0$, we have 
\ba
P_{(\lambda_1,0)}(1,t,t^2|t)=\frac{1-t^{3}}{1-t}=1+t+t^2,
\ea
when $\lambda_1>\lambda_2>0$, 
\ba
P_{(\lambda_1,\lambda_2)}(1,t,t^2|t)=t^{\lambda_2}\frac{(1-t^{3})(1-t^2)}{(1-t)(1-t)}=t^{\lambda_2}(1+t)(1+t+t^2),
\ea
and when $\lambda_1=\lambda_2=\lambda$, 
\ba
P_{(\lambda,\lambda)}(1,t,t^2|t)=t^{\lambda}\frac{(1-t^3)(1-t^2)}{(1-t)(1-t^2)}=t^\lambda(1+t+t^2).
\ea

\section{POSET Contributions in $(3,3+m)$ Model in Large $m$ Limit}\label{a:3m-POSET}

The contributions from all allowed ordering up to ${\cal O}(q^8)$ in the large $m$ limit of the vacuum module of $(3,3+m)$ model are listed in this Appendix. 
\ba
&&(0,-3,-6,-9,-12,-15,-18,\dots)\mapsto 1,\nn\\
&&(0,-m_>,-3,-6,-9,-12,-15,\dots)\mapsto Tq^2,\nn\\
&&(0,-3,-m_>,-6,-9,-12,-15,-18,\dots)\mapsto Tq^3,\nn\\
&&(0,-m,-2m_>,-3,-6,-9,-12,-15,\dots)\mapsto T^2q^3,\nn\\
&&(0,-3,-6,-m_>,-9,-12,-15,\dots)\mapsto Tq^4,\nn\\
&&(0,-3,-m,-m-3_>,-6,-9,-12,\dots)\mapsto T^2q^4,\nn\\
&&(0,-3,-m,-2m_>,-6,-9,-12,\dots)\mapsto T^2q^4,\nn\\
&&(0,-3,-6,-9,-m_>,-12,-15,\dots)\mapsto Tq^5,\nn\\
&&(0,-3,-6,-m,-m-3_>,-9,-12,\dots)\mapsto T^2q^5,\nn\\
&&(0,-3,-6,-m,-2m_>,-9,-12,\dots)\mapsto T^2q^5,\nn\\
&&(0,-3,-m,-m-3,-2m_>,-6,-9,\dots)\mapsto T^3q^5,\nn\\
&&(0,-3,-6,-9,-12,-m_>,-15,-18,\dots)\mapsto Tq^6,\nn\\
&&(0,-3,-6,-9,-m,-m-3_>,-12,-15,\dots)\mapsto T^2q^6,\nn\\
&&(0,-3,-6,-9,-m,-2m_>,-12,-15,\dots)\mapsto T^2q^6,\nn\\
&&(0,-3,-6,-m,-m-3,-m-6_>,-9,-12,\dots)\mapsto T^3q^6,\nn\\
&&(0,-3,-6,-m,-m-3,-2m_>,-9,-12,\dots)\mapsto T^3q^6,\nn\\
&&(0,-3,-m,-m-3,-2m,-2m-3_>,-6,-9,\dots)\mapsto T^4q^6,\nn\\
&&(0,-m_>,-3,-m-3_>,-6,-9,-12,\dots)\mapsto T^2q^6,\nn\\
&&(0,-m_>,-3,-2m_>,-6,-9,-12,-15,\dots)\mapsto T^3q^6,\nn\\
&&(0,-3,-6,-9,-12,-15,-m_>,-18,-21,\dots)\mapsto Tq^7,\nn\\
&&(0,-3,-6,-9,-12,-m,-m-3_>,-15,-18,\dots)\mapsto T^2q^7,\nn\\
&&(0,-3,-6,-9,-12,-m,-2m_>,-15,\dots)\mapsto T^2q^7,\nn\\
&&(0,-3,-6,-9,-m,-m-3,-m-6_>,-12,\dots)\mapsto T^3q^7,\nn\\
&&(0,-3,-6,-9,-m,-m-3,-2m_>,-12,\dots)\mapsto T^3q^7,\nn\\
&&(0,-3,-6,-m,-m-3,-m-6,-2m_>,-9,\dots)\mapsto T^4q^7,\nn
\ea
\ba
&&(0,-3,-6,-m,-m-3,-2m,-2m-3_>,-9,\dots)\mapsto T^4q^7,\nn\\
&&(0,-m_>,-3,-6,-m-3_>,-9,-12,-15,\dots)\mapsto T^2q^7,\nn\\
&&(0,-m_>,-3,-6,-2m_>,-9,\dots)\mapsto T^3q^7,\nn\\
&&(0,-m_>,-3,-m-3,-2m_>,-6,\dots)\mapsto T^3q^7,\nn\\
&&(0,-3,-m_>,-6,-m-3_>,-9,\dots)\mapsto T^2q^8,\nn\\
&&(0,-3,-m_>,-6,-2m_>,-9,\dots)\mapsto T^3q^8,\nn\\
&&(0,-m,-2m_>,-3,-m-3_>,-6,\dots)\mapsto T^4q^8,\nn\\
&&(0,-m_>,-3,-6,-9,-m-3_>,-12,\dots)\mapsto T^2q^8,\nn\\
&&(0,-m_>,-3,-6,-9,-2m_>,-12,\dots)\mapsto T^3q^8,\nn\\
&&(0,-m_>,-3,-6,-m-3,-m-6_>,-9,\dots)\mapsto T^3q^8,\nn\\
&&(0,-m_>,-3,-6,-m-3,-2m_>,-9,\dots)\mapsto T^3q^8,\nn\\
&&(0,-m_>,-3,-m-3,-2m,-2m-3_>,-6,\dots)\mapsto T^4q^8,\nn\\
&&(0,-3,-6,-9,-12,-15,-18,-m_>,-21,\dots)\mapsto Tq^8,\nn\\
&&(0,-3,-6,-9,-12,-15,-m,-m-3_>,-18,\dots)\mapsto T^2q^8,\nn\\
&&(0,-3,-6,-9,-12,-15,-m,-2m_>,-18,\dots)\mapsto T^2q^8,\nn\\
&&(0,-3,-6,-9,-12,-m,-m-3,-m-6_>,-15,\dots)\mapsto T^3q^8,\nn\\
&&(0,-3,-6,-9,-12,-m,-m-3,-2m_>,-15,\dots)\mapsto T^3q^8,\nn\\
&&(0,-3,-6,-9,-m,-m-3,-m-6,-m-9_>,-12,\dots)\mapsto T^4q^8,\nn\\
&&(0,-3,-6,-9,-m,-m-3,-m-6,-2m_>,-12,\dots)\mapsto T^4q^8,\nn\\
&&(0,-3,-6,-9,-m,-m-3,-2m,-2m-3_>,-12,\dots)\mapsto T^4q^8,\nn\\
&&(0,-3,-6,-m,-m-3,-m-6,-2m,-2m-3_>,-9,\dots)\mapsto T^5q^8.\nn\\
\ea

\section{POSET Contributions in $(3,7)$ Model}\label{a:37-POSET}

We list in this Appendix the speculated contributions from all allowed orderings in the vacuum module of $(3,7)$ model to the refined character up to ${\cal O}(q^8)$. 
\ba
&&(0,-3,-4,-6,-7,-8,-9,-10,-11,-12,-13,-14,-15,\dots)\mapsto 1,\nn\\
&&(0,-4_>,-3,-6,-7,-8,-9,-10,-11,-12,-13,-14,-15,\dots)\mapsto Tq^2,\nn\\
&&(0,-3,-6_>,-4,-7,-8,-9,-10,-11,-12,-13,-14,-15,\dots)\mapsto Tq^3,\nn\\
&&(0,-4,-8_>,-3,-6,-7,-9,-10,-11,-12,-13,-14,-15,\dots)\mapsto T^2q^3,\nn\\
&&(0,-3,-4,-7_>,-6,-8,-9,-10,-11,-12,-13,-14,-15,\dots)\mapsto Tq^4,\nn\\
&&(0,-3,-4,-8_>,-6,-7,-9,-10,-11,-12,-13,-14,-15,\dots)\mapsto T^2q^4,\nn\\
&&(0,-3,-6,-9_>,-4,-7,-8,-10,-11,-12,-13,-14,-15,\dots)\mapsto T^2q^4,\nn\\
&&(0,-3,-4,-6,-8_>,-7,-9,-10,-11,-12,-13,-14,-15,\dots)\mapsto Tq^5,\nn\\
&&(0,-3,-4,-7,-8_>,-6,-9,-10,-11,-12,-13,-14,-15,\dots)\mapsto T^2q^5,\nn\\
&&(0,-3,-4,-6,-9_>,-7,-8,-10,-11,-12,-13,-14,-15,\dots)\mapsto T^2q^5,\nn\\
&&(0,-3,-4,-6,-7,-9_>,-8,-10,-11,-12,-13,-14,-15,\dots)\mapsto Tq^6,\nn\\
&&(0,-3,-4,-6,-7,-10_>,-8,-9,-11,-12,-13,-14,-15,\dots)\mapsto T^2q^6,\nn\\
&&(0,-3,-4,-6,-8,-9_>,-7,-10,-11,-12,-13,-14,-15,\dots)\mapsto T^2q^6,\nn\\
&&(0,-3,-4,-7,-8,-11_>,-6,-9,-10,-12,-13,-14,-15,\dots)\mapsto T^3q^6,\nn\\
&&(0,-4_>,-3,-7_>,-6,-8,-9,-10,-11,-12,-13,-14,-15,\dots)\mapsto T^2q^6,\nn\\
&&(0,-4_>,-3,-8_>,-6,-7,-9,-10,-11,-12,-13,-14,-15,\dots)\mapsto T^3q^6,\nn\\
&&(0,-3,-4,-6,-7,-8,-10_>,-9,-11,-12,-13,-14,-15,\dots)\mapsto Tq^7,\nn\\
&&(0,-3,-4,-6,-7,-8,-11_>,-9,-10,-12,-13,-14,-15,\dots)\mapsto T^2q^7,\nn\\
&&(0,-3,-4,-6,-7,-9,-10_>,-8,-11,-12,-13,-14,-15,\dots)\mapsto T^2q^7,\nn\\
&&(0,-3,-4,-6,-8,-9,-12_>,-7,-10,-11,-13,-14,-15,\dots)\mapsto T^3q^7,\nn\\
&&(0,-4_>,-3,-6,-8_>,-7,-9,-10,-11,-12,-13,-14,-15,\dots)\mapsto T^2q^7,\nn\\
&&(0,-4_>,-3,-7,-8_>,-6,-9,-10,-11,-12,-13,-14,-15,\dots)\mapsto T^3q^7,\nn\\
&&(0,-4_>,-3,-6,-9_>,-7,-8,-10,-11,-12,-13,-14,-15,\dots)\mapsto T^3q^7,\nn\\
&&(0,-3,-4,-6,-7,-8,-9,-11_>,-10,-12,-13,-14,-15,\dots)\mapsto Tq^8,\nn\\
&&(0,-3,-4,-6,-7,-8,-9,-12_>,-10,-11,-13,-14,-15,\dots)\mapsto T^2q^8,\nn\\
&&(0,-3,-4,-6,-7,-8,-10,-11_>,-9,-12,-13,-14,-15,\dots)\mapsto T^2q^8,\nn
\ea
\ba
&&(0,-3,-4,-6,-7,-8,-10,-13_>,-9,-11,-12,-14,-15,\dots)\mapsto T^3q^8,\nn\\
&&(0,-3,-4,-6,-7,-9,-10,-13_>,-8,-11,-12,-14,-15,\dots)\mapsto T^3q^8,\nn\\
&&(0,-4_>,-3,-6,-7,-9_>,-8,-10,-11,-12,-13,-14,-15,\dots)\mapsto T^2q^8,\nn\\
&&(0,-4_>,-3,-6,-7,-10_>,-8,-9,-11,-12,-13,-14,-15,\dots)\mapsto T^3q^8,\nn\\
&&(0,-4_>,-3,-6,-8,-9_>,-7,-10,-11,-12,-13,-14,-15,\dots)\mapsto T^3q^8,\nn\\
&&(0,-4_>,-3,-7,-8,-11_>,-6,-9,-10,-12,-13,-14,-15,\dots)\mapsto T^4q^8,\nn\\
&&(0,-3,-6_>,-4,-8_>,-7,-9,-10,-11,-12,-13,-14,-15,\dots)\mapsto T^2q^8,\nn\\
&&(0,-4,-8_>,-3,-7_>,-6,-9,-10,-11,-12,-13,-14,-15,\dots)\mapsto T^3q^8,\nn\\
\ea

\bibliography{Macdonald-index}

\end{document}